\documentclass[twocolumn]{revtex4}

\usepackage{amsfonts}
\usepackage{amsmath}
\usepackage{amssymb}
\usepackage{ragged2e}
\usepackage{graphicx}
\usepackage{bm}

\usepackage[usenames,dvipsnames]{color}
\definecolor{darkblue}{RGB}{0,0,196}

\begin{document}
\title{Analysis of identified particle transverse momentum spectra produced in pp, p--Pb and Pb--Pb
collisions at the LHC using Tsallis--Pareto-type distribution
\vspace{0.5cm}}

\author{Pei-Pin~Yang$^{1,}$\footnote{E-mails: peipinyangshanxi@163.com; yangpeipin@qq.com},
Mai-Ying~Duan$^{1,}$\footnote{E-mail: duanmaiying@sxu.edu.cn},
Fu-Hu~Liu$^{1,}$\footnote{Correspondence: E-mails:
fuhuliu@163.com; fuhuliu@sxu.edu.cn},
Raghunath~Sahoo$^{2,3,}$\footnote{Correspondence: E-mails:
Raghunath.Sahoo@cern.ch; raghunath@iiti.ac.in}}

\affiliation{$^1$Institute of Theoretical Physics \& College of
Physics and Electronic Engineering, State Key Laboratory of
Quantum Optics and Quantum Optics Devices \& Collaborative
Innovation Center of Extreme Optics, Shanxi University, Taiyuan
030006, China
\\
$^2$Department of Physics, Indian Institute of Technology Indore,
Indore 453552, India
\\
$^3$Experimental Physics Department, CERN, 1211 Geneva 23,
Switzerland}

\begin{abstract}

\vspace{0.5cm}

\noindent {\bf Abstract:} In the framework of a multi-source
thermal model at the quark level, we have analyzed transverse
momentum spectra of hadrons measured by the ALICE Collaboration in
proton--proton ($pp$ or $p$--$p$) collisions at the center-of-mass
energy of $\sqrt{s}=7$ and 13 TeV, proton--lead ($p$--Pb)
collisions at $\sqrt{s_{\rm NN}}=5.02$ TeV, and lead--lead
(Pb--Pb) collisions at $\sqrt{s_{\rm NN}}=2.76$ TeV. For
meson(baryon), the contributions of two(three) constituent quarks
are considered, in which each quark contributes to hadron
transverse momentum to obey the revised Tsallis--Pareto-type
distribution (the TP-like function in short) with isotropic random
azimuth. Three main parameters, namely, the revised index $a_0$,
effective temperature $T$, and entropy-related index $n$ are
obtained, which show the same tendency for small and large systems
with respect to the centrality (or multiplicity) of events, rest
mass of hadrons, and constituent mass of quarks.
\\
\\
{\bf Keywords:} Transverse momentum spectra, Tsallis--Pareto
distribution, constituent quarks, proton--proton and heavy-ion
Collisions
\\
\\
{\bf PACS numbers:} 12.40.Ee, 13.85.Hd, 24.10.Pa
\end{abstract}

\maketitle

\section{Introduction}

The deconfined high-temperature and high-density state of nuclear
matter, called quark-gluon plasma (QGP) is often studied by
ultra-relativistic nuclear collisions. The transverse momentum
($p_T$) spectra and ratios of identified hadrons provide a means
to study the properties of matter created in these collisions and
the mechanism by which quasi-free partons are transformed into
observable hadrons. The relative contributions of different
hadronization mechanisms rely on the change of hadron's $p_T$. At
low-$p_T$, recombination may be dominant, but at high-$p_T$,
hadrons may originate from fragmentation processes. This mainly
depends on the potential transverse momentum distribution of
quarks. Therefore, it is important to have the $p_T$-spectra of
identified mesons and baryons in a wide $p_T$-range.

Unlike up ($u$) and down ($d$) quarks those form ordinary matter,
strange ($s$) quarks do not exist in the form of valence quarks in
the colliding species, but they are light enough to be produced in
large quantities in the process of ultra-relativistic collisions.
In the early state of high-energy collisions, strangeness is
produced in hard (perturbative) $2\rightarrow2$ partonic
scattering processes by flavor creation ($gg\rightarrow s\bar{s}$,
$q\bar{q}\rightarrow s\bar{s}$) and flavor excitation
($gs\rightarrow gs$, $qs\rightarrow qs$). Strangeness is also
created during the subsequent partonic evolution via gluon
splitting process ($g\rightarrow s\bar{s}$). These processes tend
to dominate the production of strange hadrons with high-$p_T$, and
the production of strange hadrons with low-$p_T$ is mainly
dominated by non-perturbative processes.

The production of strange hadrons is suppressed compared to that
of hadrons containing only $u/d$ quarks, because $s$ quarks need
higher threshold energy to be excited. In basic particle-particle
collisions, the degree of suppression of strange hadrons is an
important parameter in the model analysis. Therefore, the
measurement of strange hadron production imposes restrictions on
the models. The study of strange and multi-strange particles in
relativistic heavy-ion collisions is an important tool for
exploring the properties of the strong interaction system. The
particle spectra provide abundant information about the
temperature and collective flow of the system, which reflects the
dynamic freezing condition, where the inelastic collision stops.
The enhancement of strangeness in heavy-ion collisions is one of
the most important signals of QGP~\cite{1,2,3}.

In the past decades, around the search for QGP, researchers have
conducted extensive study on hadrons containing one or more $s$
quarks~\cite{1,4,5}. However, the origin of the strangeness
enhancement is not yet clear, when strangeness is also observed to
be enhanced in proton--proton ($pp$ or $p$--$p$)
collisions~\cite{6,7}. The azimuthal correlations and
mass-dependent hardening of the $p_T$ spectra observed in the
high-multiplicity $pp$ and proton--nucleus ($pA$ or $p$--$A$)
collisions are typically attributed to the formation of strongly
interacting quark-gluon
medium~\cite{8,9,10,11,12,13,14,15,16,17,18,19}. The abundance of
strange particles at different center-of-mass energies is in
accordance with the calculation of the thermal statistical
model~\cite{20,21,22}. Strangeness, light flavor production, and
heavy-ion collision dynamics provide evidences for the properties
of fluid-like and collectiveness of the medium~\cite{23,24}.
Studying $pp$ collisions under high multiplicity is of
considerable significance, because it opens up the possibility of
understanding nuclear reaction phenomena from a microscopic
perspective.

Studies of high multiplicity charged particles in $pp$ and
proton--lead ($p$--Pb) collisions at the Large Hadron Collider
(LHC) have shown striking similarities to lead--lead (Pb--Pb)
collisions. The facts that the enhancements of (multi-)strange
hadrons~\cite{11}, azimuthal correlations and double-ridge
structure~\cite{12,13}, nonzero elliptic flow ($v_2$) coefficients
and other anisotropic flow measurements~\cite{25,26,27,28,29,30},
mass ordering in hadron $p_T$ spectra, and characteristic
modifications of baryon to meson ratios~\cite{10}, have shown that
collective phenomena also exist in small collision systems. In
addition, the continuous evolution of the ratio of light-flavor
hadrons to pions in $pp$, $p$--Pb, and Pb--Pb collisions have been
found as a function of charged particle multiplicity
density~\cite{26,31,32}. The observed similarities indicate that
there is a common underlying mechanism that determines the
chemical composition that arises from these collisions. That is,
typically interactions among partons exit in these collisions.

Recently, we considered contributor quarks and based on the
Tsallis statistics~\cite{33,34,35,36,37}, the available $p_T$
spectra of various particles produced in collisions of small
systems [$pp$, deuteron--gold ($d$--Au), $p$--Pb] and large
systems [gold--gold (Au--Au) and Pb--Pb] at high energies were
studied~\cite{38,39}, where we have used the convolution of two or
three revised Tsallis--Pareto-type distribution (the TP-like
function in short)~\cite{40,41,42,43}. The application of
convolution means that we have considered the azimuths of
contributor quarks to be the same or parallel to each other. A
detailed consideration shows that the azimuths of contributor
quarks may be isotropic and random which includes the parallel,
vertical, and any other cases. The analytical expressions for
parallel and vertical cases are available. For any other cases, we
may use the Monte Carlo method, if it is difficult to give the
analytical expression.

In the current work, in the framework of multi-source thermal
model at the quark level, we use the Monte Carlo method to
study the $p_T$ spectra of identified hadrons produced in the
collisions of different centralities (or multiplicities) with
small systems ($pp$ and $p$--Pb) and large system (Pb--Pb),
including non-strange hadrons ($\pi^++\pi^-$ and
$p+\overline{p}$), strange hadrons ($K^++K^-$, $K_S^0$,
$K^*+\overline{K}{}^*$, and $\Lambda+\overline\Lambda$), and
multi-strange hadrons ($\Xi^-+\overline\Xi{}^+$ and
$\Omega^-+\overline\Omega{}^+$). The azimuths of contributor or
constituent quarks are isotropic and random. The size of
transverse momentum of each quark contributed to hadron's $p_T$ is
assumed to obey the TP-like function. Mathematically, we
study the synthesis of two or three vectors with changeable
azimuths and sizes.

The $p_T$ spectra in a wide range can reflect more dynamical
information of the collision process. In order to verify the
feasibility of the model and extract relevant parameters, we
collected the experimental data of $p_T$ spectra of identified
hadrons produced in $pp$ collisions at $\sqrt{s}=7$ and 13
TeV~\cite{44,45,46,47}, $p$--Pb collisions at $\sqrt{s_{\rm
NN}}=5.02$ TeV~\cite{10,48,49,50}, and Pb--Pb collisions at
$\sqrt{s_{\rm NN}}=2.76$ TeV~\cite{51,52,53,54}, measured by the
ALICE Collaboration.

The remainder of this paper is structured as follows. The
formalism and method are described in Section II. Results and
discussion are given in Section III. In Section IV, we summarize our
main observations with conclusions.

\section{Formalism and method}

Inspired by the SU(3) super polymorphism theory, in 1964, M.
Gell-Mann proposed the quark model~\cite{55}. There are two
commonly used masses of quarks. One is the current mass of quark,
which refers to the mass in the Lagrangian of quantum field
theory. The other is the constituent mass of quark that refers to
the equivalent mass after the interaction with the gluon being
included when considering the composition of hadrons. The quark
model believes that a meson is composed of a quark and antiquark
pair, and a baryon is composed of three quarks.

In our recent work~\cite{38,39}, the TP-like function used for the
$p_T$ spectra of hadrons is given by
\begin{align}
f(p_{T})=\frac{1}{N}\frac{dN}{dp_T}=
C_0p_{T}^{a_0}\left(1+\frac{\sqrt{p_{T}^2+m_0^2}-m_0}{nT}
\right)^{-n},
\end{align}
where $N$ is the number of particles, $C_0$ is the normalization
constant, $a_{0}$ is the revised factor, $T$ is the effective
temperature, $n$ is the entropy-related index, and $m_0$ is the
rest mass of given hadron. The three parameters $a_{0}$, $T$, and
$n$ can be determined from the $p_T$ spectra of hadrons.

According to the multi-source thermal model~\cite{56,57}, we think
that the $p_T$ spectra of hadrons are contributed by the
contributor or constituent quarks. The transverse momentum of each
quark contributed to $p_T$ of given hadron is assumed to follow
the TP-like function, too. We have the contribution, $p_{ti}$, of
the $i$-th quark to be
\begin{align}
f_i(p_{ti})=\frac{1}{N_i}\frac{dN_i}{dp_T}= C_ip_{ti}^{a_0} \left(
1+ \frac{\sqrt{p_{ti}^2+m_{0i}^2}-m_{0i}}{nT} \right)^{-n}.
\end{align}
Here, $N_i$ is number of the $i$-th quark, $C_i$ is the
normalization constant, and $m_{0i}$ is the constituent mass of
the $i$-th quark. A meson is composed of a quark and antiquark pair, so $i$ is
equal to 1 or 2, and a baryon is composed of 3 quarks, so $i$ is
equal to 1, 2 or 3. Regardless of the value of $i$, we always have
$N_i=N$.

The relations between the transverse momentum vectors
$\bm{p_{t1}}$ and $\bm{p_{t2}}$ of constituent quark and antiquark pair for meson,
and those among the transverse momentum vectors $\bm{p_{t1}}$,
$\bm{p_{t2}}$, and $\bm{p_{t3}}$ of constituent quarks for baryon,
may be parallel, vertical, or of any azimuth $\phi_i$. The analytical
expressions for the parallel and vertical cases are
available in Refs.~\cite{38,39}. For the cases of any azimuths, we may use
the Monte Carlo method~\cite{58,59} to obtain $p_T$, if the
analytical expression is not available.

To obtain a discrete value of $p_{ti}$ that satisfies Eq. (2), we
can perform the solution of
\begin{align}
\int_{0}^{p_{ti}}f_{i}(p'_{ti})dp'_{ti}<r_{i}<\int_{0}^{p_{ti}+\delta
p_{ti}}f_{i}(p'_{ti})dp'_{ti}.
\end{align}
Here, $r_{i}$ is a random number uniformly distributed in $[0,1]$,
$\delta p_{ti}$ denotes a small shift in $p_{ti}$. To obtain a
discrete value of $\phi_i$ that satisfies the isotropic or uniform
distribution, we have
\begin{align}
\phi_i=2\pi R_i,
\end{align}
where $R_i$ denotes a random number uniformly distributed in
$[0,1]$.

For a meson composed of a quark and antiquark pair, we have,
\begin{align}
p_T=&\sqrt{\bigg(\sum_{i=1}^2 p_{ti}\cos\phi_i\bigg)^2
+\bigg(\sum_{i=1}^2 p_{ti}\sin\phi_i\bigg)^{2}}\nonumber\\
=&\sqrt{p_{t1}^{2}+p_{t2}^{2}+2p_{t1}p_{t2}\cos|\phi_{1}-\phi_{2}|}.
\end{align}
For a baryon composed of three quarks, we have,
\begin{align}
p_T=&\sqrt{\bigg(\sum_{i=1}^3 p_{ti}\cos\phi_i\bigg)^2
+\bigg(\sum_{i=1}^3 p_{ti}\sin\phi_i\bigg)^{2}}\nonumber\\
=&\big(p_{t1}^{2}+p_{t2}^{2}+p_{t3}^{2}+2p_{t1}p_{t2}\cos|\phi_{1}-\phi_{2}|\nonumber\\
&+2p_{t1}p_{t3}\cos|\phi_{1}-\phi_{3}|+2p_{t2}p_{t3}\cos|\phi_{2}-\phi_{3}|\big)^{1/2}.
\end{align}

As the expression of $p_T$ for hadron with 2 or 3 quarks, Eq. (5)
or (6) can be easily extended to the hadron with multiple quarks,
e.g. the hadron with 4 or 5 quarks. We have the expression of
$p_T$ for hadron with $k$ quarks to be
\begin{align}
p_T=\sqrt{\bigg(\sum_{i=1}^k p_{ti}\cos\phi_i\bigg)^2
+\bigg(\sum_{i=1}^k p_{ti}\sin\phi_i\bigg)^{2}}.
\end{align}
In fact, Eq. (7) is a uniform expression for $p_T$ of any hadron.
In a real calculation, we need $k$ sets of $p_{ti}$ and $\phi_i$
for a given $p_T$. After the calculation for many times, the
distribution of $p_T$ can be obtained in statistics.

It is worth emphasized that in the analysis performed in this
paper, the mass of the quark is the constituent mass, but not the
current mass~\cite{60a,60}. Our attempts show that the constituent
mass is more suitable for the fit. Meanwhile, we need to always
distinguish the transverse momentum, $p_T$ of a given hadron and
the transverse momentum $p_{ti}$ of the $i$-th constituent quark
with the azimuth $\phi_i$.

\begin{figure*}[htbp]
\begin{center}
\includegraphics[width=11.0cm]{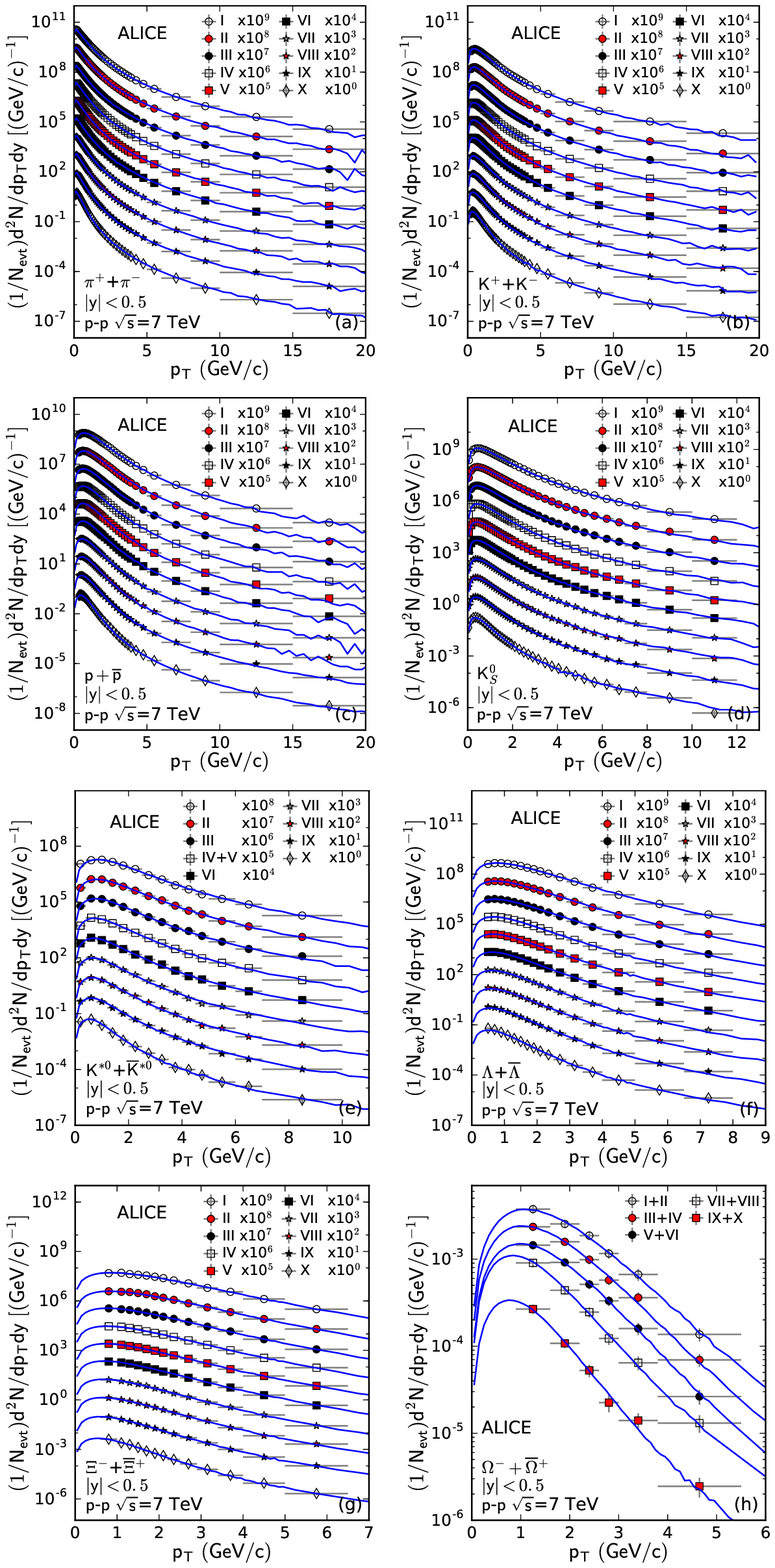}
\end{center}
\justifying\noindent {Fig. 1. Multiplicity dependent $p_{T}$
spectra of $\pi^++\pi^-$ (a), $K^{+}+K^{-}$ (b), $p+\overline{p}$
(c), $K^{0}_{S}$ (d), $K^{*}+\overline{K}{}^*$ (e),
$\Lambda+\overline{\Lambda}$ (f), $\Xi^{-}+\overline\Xi{}^{+}$
(g), and $\Omega^{-}+\overline{\Omega}{}^{+}$ (h) with $|y|<0.5$,
produced in $pp$ collisions at $\sqrt{s}=7$ TeV. Different symbols
represent the experimental data for different multiplicity classes
measured by the ALICE Collaboration~\cite{44}, where in most cases
the data are scaled by constant multipliers marked in the panels
for clarity. The curves represent our fit results based on the
Monte Carlo calculations.}
\end{figure*}

\begin{figure*}[htbp]
\begin{center}
\includegraphics[width=11.0cm]{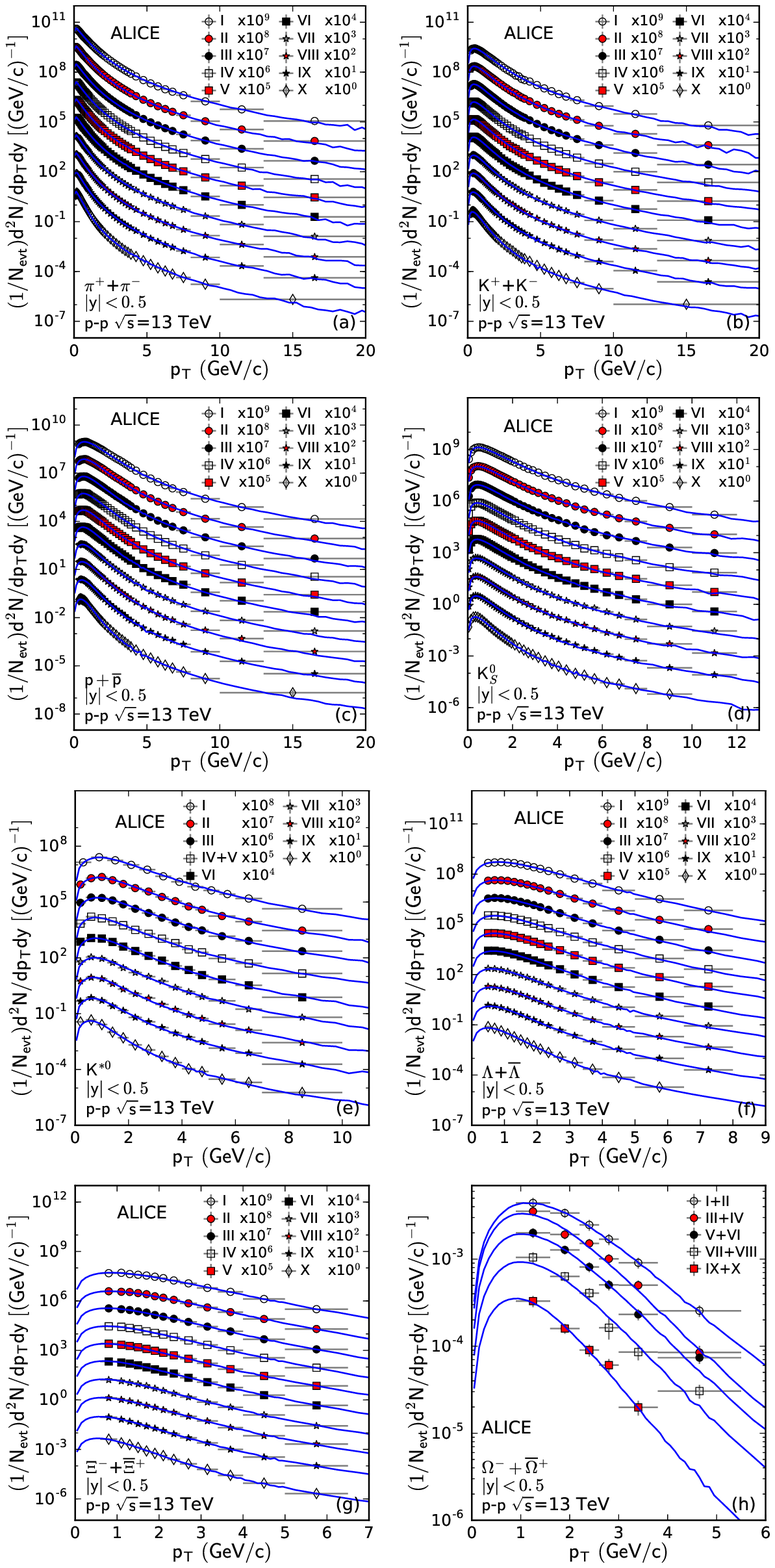}
\end{center}
\justifying\noindent {Fig. 2. Same as Figure 1, but showing the
multiplicity dependent $p_{T}$ spectra of $\pi^++\pi^-$ (a),
$K^{+}+K^{-}$ (b), $p+\overline{p}$ (c), $K^{0}_{S}$ (d), $K^{*0}$
(e), $\Lambda+\overline{\Lambda}$ (f),
$\Xi^{-}+\overline\Xi{}^{+}$ (g), and
$\Omega^{-}+\overline{\Omega}{}^{+}$ (h) with $|y|<0.5$ produced
in $pp$ collisions at $\sqrt{s}=13$ TeV measured by the ALICE
Collaboration~\cite{45,46,47}.}
\end{figure*}

\begin{figure*}[htbp]
\begin{center}
\includegraphics[width=11.0cm]{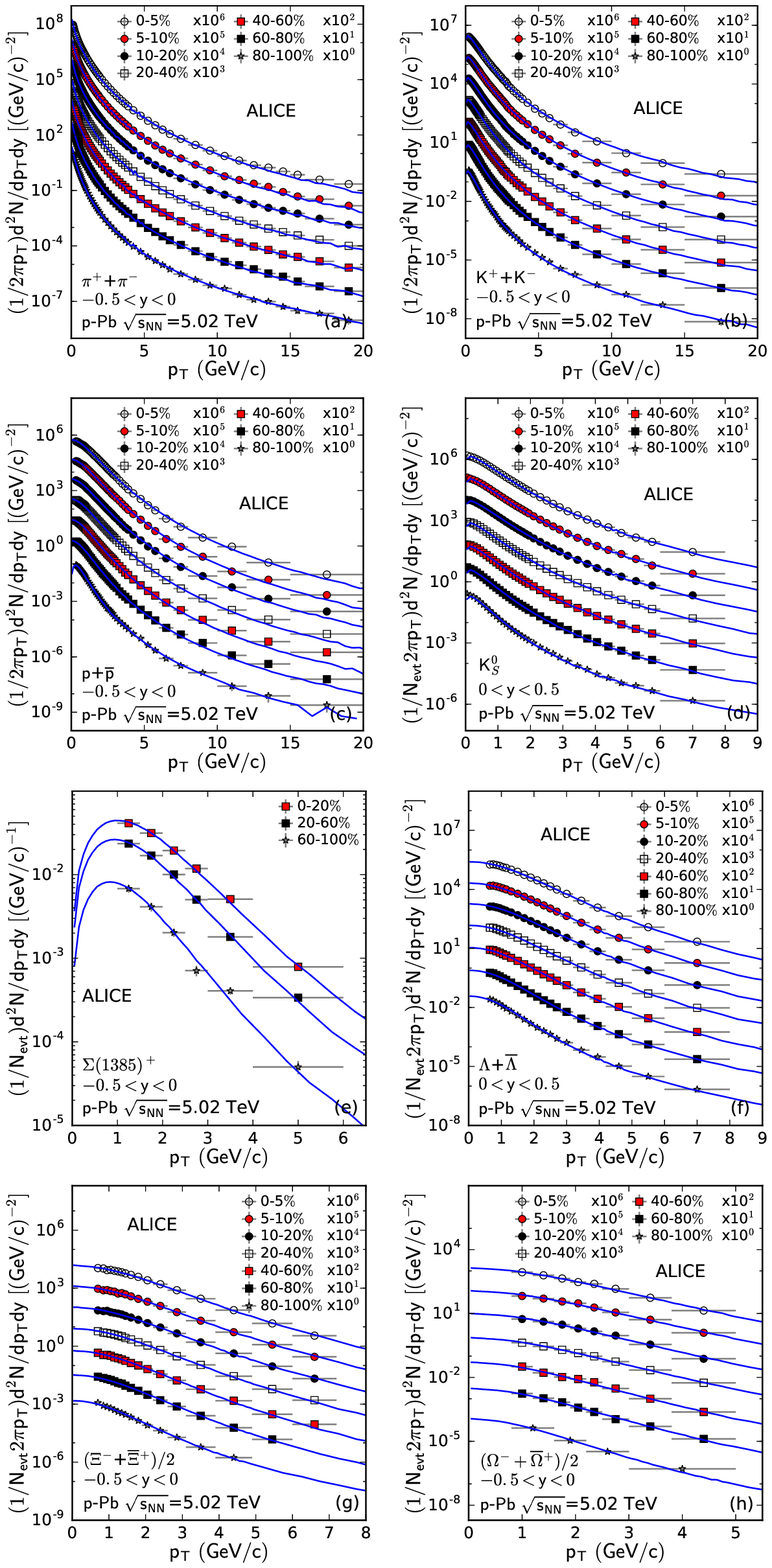}
\end{center}
\justifying\noindent {Fig. 3. The centrality dependent $p_T$
spectra of $\pi^++\pi^-$ (a), $K^{+}+K^{-}$ (b), $p+\overline{p}$
(c), $K^{0}_{S}$ (d), $\Sigma(1385)^+$ (e),
$\Lambda+\overline{\Lambda}$ (f), $(\Xi^{-}+\overline\Xi{}^{+})/2$
(g), and $(\Omega^{-}+\overline{\Omega}{}^{+})/2$ (h) with
$-0.5<y<0$ (a--c, e, g, h) or $0<y<0.5$ (d, f) produced in $p$--Pb
collisions at $\sqrt{s_{NN}}=5.02$ TeV measured by the ALICE
Collaboration~\cite{10,48,49,50}. Different symbols represent the
experimental data with different centrality classes, where
different constant multipliers are used to re-scale the data for clarity. The
curves are our fit results based on the Monte Carlo calculations.}
\end{figure*}

\begin{figure*}[htbp]
\begin{center}
\includegraphics[width=11.0cm]{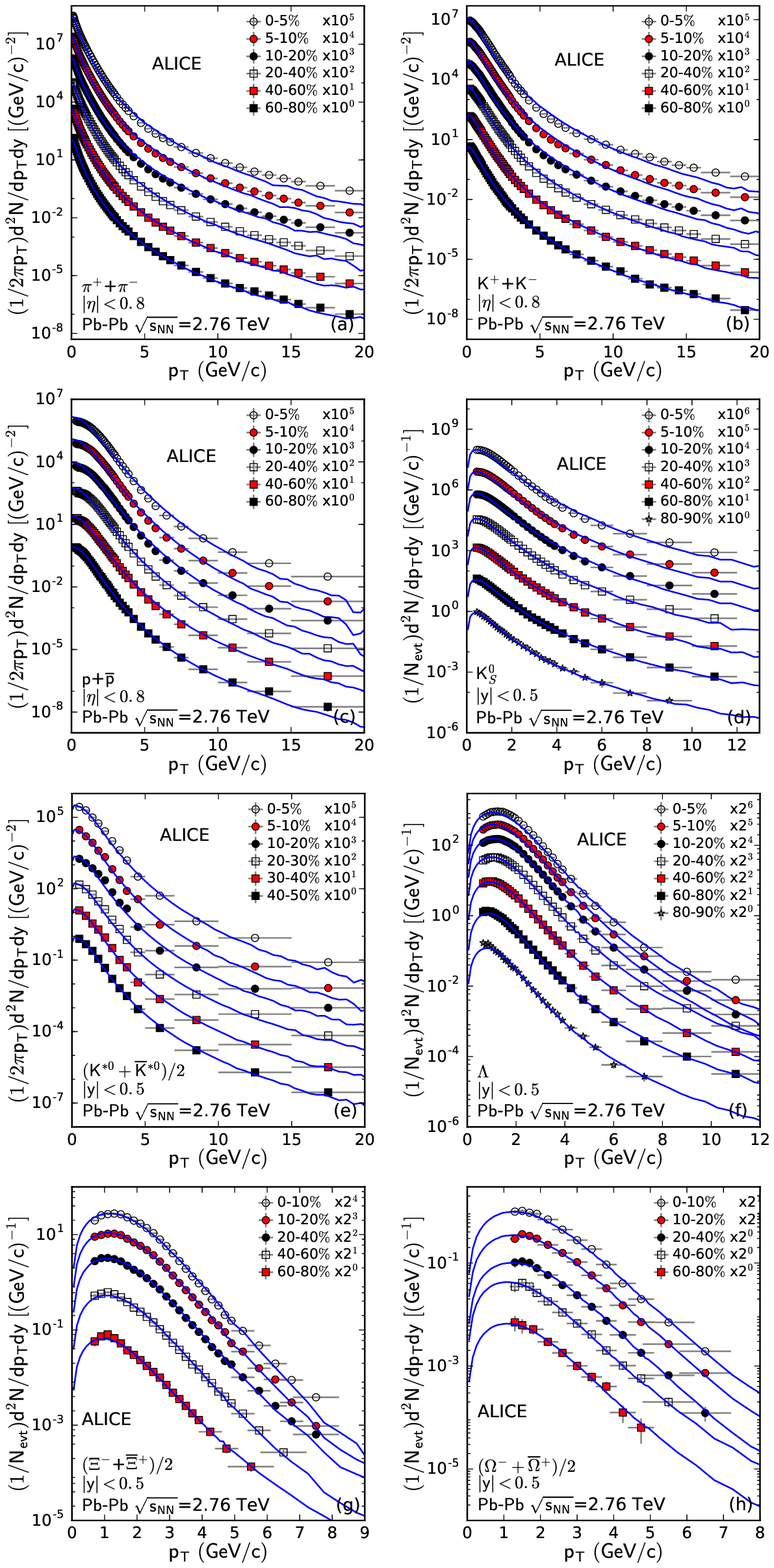}
\end{center}
\justifying\noindent {Fig. 4. Same as Figure 3, but showing the
centrality dependent $p_T$ spectra, of $\pi^++\pi^-$ (a),
$K^{+}+K^{-}$ (b), $p+\overline{p}$ (c), $K^0_S$ (d),
$(K^{*0}+\overline{K}{}^{*0})/2$ (e), $\Lambda$ (f),
$(\Xi^{-}+\overline{\Xi}{}^{+})/2$ (g), and
$(\Omega^{-}+\overline{\Omega}{}^{+})/2$ (h) with $|\eta|<0.8$
(a--c) or $|y|<0.5$ (d--h) produced in Pb--Pb collisions at
$\sqrt{s_{NN}}=2.76$ TeV measured by the ALICE
Collaboration~\cite{51,52,53,54}.}
\end{figure*}

\begin{table*}[htbp]\vspace{0.2cm}
\justifying\noindent {\small Table 1. Values of the free
parameters ($a_{0}$, $T$, and $n$), normalization constant
($N_0$), $\chi^2$, and ndof corresponding to the curves in Figure
1 for $pp$ collisions at $\sqrt{s}=7$ TeV. The particle type
(quark structure), spectrum form, and multiplicity classes are
given together. The multiplicity classes I, II, III, IV, V, VI,
VII, VIII, IX, and X correspond to the centrality classes
0--0.95\%, 0.95--4.7\%, 4.7--9.5\%, 9.5--14\%, 14--19\%, 19--28\%,
28--38\%, 38--48\%, 48--68\%, and 68--100\%, respectively.

\vspace{-0.35cm}

\begin{center}
\newcommand{\tabincell}[2]{\begin{tabular}{@{}#1@{}}#2\end{tabular}}
\begin{tabular} {cccccccccccc}\\ \hline\hline
\tabincell{c}{Particle (quark structure)\\ and spectrum form}& \tabincell{c}{Multiplicity\\ class} & $a_{0}$ & $T$ (GeV) & $n$ & $N_0$ & $\chi^2$/ndof \\
\hline
                              & I    & $-0.550\pm0.004$ & $0.345\pm0.002$ & $5.071\pm0.054$ & $(2.130\pm0.042)\times10^{1}$ & $58/44$\\
                              & II   & $-0.550\pm0.003$ & $0.321\pm0.003$ & $4.962\pm0.035$ & $(1.700\pm0.021)\times10^{1}$ & $66/44$\\
                              & III  & $-0.550\pm0.004$ & $0.313\pm0.002$ & $4.962\pm0.031$ & $(1.350\pm0.040)\times10^{1}$ & $69/44$\\
$\pi^{+}+\pi^{-}$             & IV   & $-0.550\pm0.004$ & $0.299\pm0.002$ & $4.934\pm0.066$ & $(1.200\pm0.030)\times10^{1}$ & $66/44$\\
$(u\bar{d},d\bar{u})$         & V    & $-0.550\pm0.005$ & $0.291\pm0.001$ & $4.910\pm0.033$ & $(1.040\pm0.101)\times10^{1}$ & $65/44$\\
$(1/N_{evt}) d^2N/dp_{T}dy$   & VI   & $-0.550\pm0.005$ & $0.287\pm0.001$ & $4.906\pm0.029$ & $(8.580\pm0.080)\times10^{0}$ & $84/44$\\
$[({\rm GeV}/c)^{-1}]$        & VII  & $-0.550\pm0.004$ & $0.266\pm0.002$ & $4.910\pm0.086$ & $(7.230\pm0.040)\times10^{0}$ & $62/44$\\
                              & VIII & $-0.550\pm0.003$ & $0.254\pm0.002$ & $4.910\pm0.115$ & $(5.860\pm0.030)\times10^{0}$ & $69/44$\\
                              & IX   & $-0.550\pm0.003$ & $0.233\pm0.001$ & $4.916\pm0.041$ & $(4.370\pm0.101)\times10^{0}$ & $104/44$\\
                              & X    & $-0.550\pm0.005$ & $0.187\pm0.002$ & $4.910\pm0.032$ & $(2.750\pm0.080)\times10^{0}$ & $194/44$\\
\hline
                              & I    & $0.148\pm0.010$ & $0.260\pm0.002$ & $5.325\pm0.028$ & $(2.810\pm0.080)\times10^{0}$  & $23/39$\\
                              & II   & $0.148\pm0.012$ & $0.248\pm0.001$ & $5.321\pm0.026$ & $(2.140\pm0.040)\times10^{0}$  & $11/39$\\
                              & III  & $0.148\pm0.024$ & $0.237\pm0.002$ & $5.260\pm0.018$ & $(1.740\pm0.020)\times10^{0}$  & $7/39$\\
$K^{+}+K^{-}$                 & IV   & $0.148\pm0.016$ & $0.230\pm0.001$ & $5.260\pm0.014$ & $(1.490\pm0.020)\times10^{0}$  & $6/39$\\
$(u\bar{s},s\bar{u})$         & V    & $0.148\pm0.016$ & $0.224\pm0.001$ & $5.260\pm0.014$ & $(1.300\pm0.010)\times10^{0}$  & $5/39$\\
$(1/N_{evt}) d^2N/dp_{T}dy$   & VI   & $0.148\pm0.020$ & $0.216\pm0.001$ & $5.261\pm0.018$ & $(1.090\pm0.020)\times10^{0}$  & $10/39$\\
$[({\rm GeV}/c)^{-1}]$        & VII  & $0.148\pm0.016$ & $0.207\pm0.001$ & $5.252\pm0.023$ & $(8.500\pm0.080)\times10^{-1}$ & $11/39$\\
                              & VIII & $0.148\pm0.018$ & $0.194\pm0.001$ & $5.240\pm0.025$ & $(6.860\pm0.100)\times10^{-1}$ & $17/39$\\
                              & IX   & $0.148\pm0.016$ & $0.178\pm0.001$ & $5.214\pm0.029$ & $(5.000\pm0.060)\times10^{-1}$ & $22/39$\\
                              & X    & $0.148\pm0.021$ & $0.146\pm0.002$ & $5.220\pm0.030$ & $(2.900\pm0.060)\times10^{-1}$ & $34/39$\\
\hline
                              & I    & $0.530\pm0.008$ & $0.214\pm0.002$ & $6.645\pm0.035$ & $(1.109\pm0.009)\times10^{0}$  & $5/37$\\
                              & II   & $0.530\pm0.005$ & $0.205\pm0.002$ & $6.646\pm0.075$ & $(8.701\pm0.150)\times10^{-1}$ & $20/37$\\
                              & III  & $0.530\pm0.008$ & $0.194\pm0.001$ & $6.579\pm0.063$ & $(7.079\pm0.040)\times10^{-1}$ & $17/37$\\
$p+\overline{p}$              & IV   & $0.530\pm0.009$ & $0.185\pm0.001$ & $6.603\pm0.046$ & $(6.130\pm0.080)\times10^{-1}$ & $14/37$\\
$(uud,\bar{u}\bar{u}\bar{d})$ & V    & $0.530\pm0.008$ & $0.178\pm0.003$ & $6.580\pm0.086$ & $(5.450\pm0.090)\times10^{-1}$ & $23/37$\\
$(1/N_{evt}) d^2N/dp_{T}dy$   & VI   & $0.530\pm0.011$ & $0.170\pm0.001$ & $6.480\pm0.080$ & $(4.530\pm0.080)\times10^{-1}$ & $24/37$\\
$[({\rm GeV}/c)^{-1}]$        & VII  & $0.530\pm0.004$ & $0.157\pm0.002$ & $6.353\pm0.066$ & $(3.650\pm0.060)\times10^{-1}$ & $16/37$\\
                              & VIII & $0.530\pm0.009$ & $0.147\pm0.001$ & $6.341\pm0.077$ & $(2.900\pm0.050)\times10^{-1}$ & $22/37$\\
                              & IX   & $0.530\pm0.008$ & $0.134\pm0.001$ & $6.181\pm0.076$ & $(2.079\pm0.020)\times10^{-1}$ & $25/37$\\
                              & X    & $0.530\pm0.010$ & $0.110\pm0.001$ & $6.180\pm0.065$ & $(1.120\pm0.020)\times10^{-1}$ & $21/37$\\
\hline
                              & I    & $0.148\pm0.008$ & $0.281\pm0.002$ & $5.722\pm0.046$ & $(1.338\pm0.015)\times10^{0}$  & $10/34$\\
                              & II   & $0.148\pm0.007$ & $0.269\pm0.002$ & $5.720\pm0.057$ & $(1.030\pm0.010)\times10^{0}$  & $35/34$\\
                              & III  & $0.148\pm0.008$ & $0.257\pm0.001$ & $5.716\pm0.045$ & $(8.360\pm0.010)\times10^{-1}$ & $29/34$\\
$K^{0}_{S}$                   & IV   & $0.148\pm0.009$ & $0.247\pm0.002$ & $5.625\pm0.057$ & $(7.080\pm0.080)\times10^{-1}$ & $35/34$\\
$(d\bar{s})$                  & V    & $0.148\pm0.008$ & $0.239\pm0.002$ & $5.560\pm0.057$ & $(6.200\pm0.080)\times10^{-1}$ & $34/34$\\
$(1/N_{evt}) d^2N/dp_{T}dy$   & VI   & $0.148\pm0.011$ & $0.231\pm0.001$ & $5.545\pm0.066$ & $(5.150\pm0.060)\times10^{-1}$ & $46/34$\\
$[({\rm GeV}/c)^{-1}]$        & VII  & $0.148\pm0.008$ & $0.222\pm0.001$ & $5.538\pm0.088$ & $(4.000\pm0.040)\times10^{-1}$ & $70/34$\\
                              & VIII & $0.148\pm0.010$ & $0.210\pm0.001$ & $5.480\pm0.118$ & $(3.220\pm0.040)\times10^{-1}$ & $70/34$\\
                              & IX   & $0.148\pm0.007$ & $0.194\pm0.001$ & $5.444\pm0.096$ & $(2.350\pm0.030)\times10^{-1}$ & $63/34$\\
                              & X    & $0.148\pm0.011$ & $0.154\pm0.001$ & $5.200\pm0.057$ & $(1.270\pm0.020)\times10^{-1}$ & $35/34$\\
\hline
\end{tabular}%
\end{center}}
\end{table*}

\begin{table*}[htb!]\vspace{0.2cm}
\justifying\noindent {\small Table 1. Continued. \vspace{-0.35cm}
\begin{center}
\newcommand{\tabincell}[2]{\begin{tabular}{@{}#1@{}}#2\end{tabular}}
\begin{tabular} {cccccccccccc}\\ \hline\hline
\tabincell{c}{Particle (quark structure)\\ and spectrum form}& \tabincell{c}{Multiplicity\\ class} & $a_{0}$ & $T$ (GeV) & $n$ & $N_0$ & $\chi^2$/ndof \\
\hline
                              & I    & $1.450\pm0.012$ & $0.196\pm0.001$ & $7.073\pm0.030$ & $(3.250\pm0.040)\times10^{-1}$ & $5/10$\\
                              & II   & $1.450\pm0.006$ & $0.186\pm0.001$ & $7.062\pm0.039$ & $(2.700\pm0.029)\times10^{-1}$ & $3/10$\\
$K^{*0}+\overline{K}{}^{*0}$  & III  & $1.450\pm0.013$ & $0.168\pm0.002$ & $6.780\pm0.035$ & $(2.300\pm0.040)\times10^{-1}$ & $5/10$\\
$(u\bar{s},s\bar{u})$         & IV+V & $1.450\pm0.018$ & $0.154\pm0.002$ & $6.569\pm0.035$ & $(1.910\pm0.030)\times10^{-1}$ & $3/10$\\
$(1/N_{evt}) d^2N/dp_{T}dy$   & VI   & $1.450\pm0.020$ & $0.141\pm0.002$ & $6.376\pm0.086$ & $(4.519\pm0.020)\times10^{-1}$ & $9/10$\\
$[({\rm GeV}/c)^{-1}]$        & VII  & $1.450\pm0.012$ & $0.133\pm0.002$ & $6.365\pm0.057$ & $(1.249\pm0.040)\times10^{-1}$ & $9/10$\\
                              & VIII & $1.450\pm0.016$ & $0.124\pm0.003$ & $6.367\pm0.066$ & $(1.039\pm0.020)\times10^{-1}$ & $6/10$\\
                              & IX   & $1.450\pm0.014$ & $0.114\pm0.002$ & $6.346\pm0.108$ & $(7.800\pm0.040)\times10^{-2}$ & $10/10$\\
                              & X    & $1.450\pm0.020$ & $0.093\pm0.002$ & $6.315\pm0.111$ & $(4.800\pm0.030)\times10^{-2}$ & $10/10$\\
\hline
                              & I    & $1.100\pm0.004$ & $0.177\pm0.001$ & $7.149\pm0.021$ & $(7.619\pm0.089)\times10^{-1}$ & $3/12$\\
                              & II   & $1.100\pm0.004$ & $0.166\pm0.001$ & $7.138\pm0.036$ & $(5.900\pm0.070)\times10^{-1}$ & $7/12$\\
                              & III  & $1.100\pm0.006$ & $0.158\pm0.002$ & $7.078\pm0.038$ & $(4.700\pm0.060)\times10^{-1}$ & $12/12$\\
$\Lambda+\overline{\Lambda}$  & IV   & $1.100\pm0.015$ & $0.151\pm0.002$ & $6.996\pm0.040$ & $(3.979\pm0.060)\times10^{-1}$ & $16/12$\\
$(uds,\bar{u}\bar{d}\bar{s})$ & V    & $1.100\pm0.013$ & $0.144\pm0.002$ & $6.868\pm0.042$ & $(3.450\pm0.060)\times10^{-1}$ & $16/12$\\
$(1/N_{evt}) d^2N/dp_{T}dy$   & VI   & $1.100\pm0.022$ & $0.136\pm0.002$ & $6.780\pm0.061$ & $(2.850\pm0.040)\times10^{-1}$ & $18/12$\\
$[({\rm GeV}/c)^{-1}]$        & VII  & $1.100\pm0.014$ & $0.126\pm0.002$ & $6.526\pm0.075$ & $(2.200\pm0.039)\times10^{-1}$ & $21/12$\\
                              & VIII & $1.100\pm0.011$ & $0.118\pm0.003$ & $6.470\pm0.103$ & $(1.720\pm0.040)\times10^{-1}$ & $27/12$\\
                              & IX   & $1.100\pm0.012$ & $0.111\pm0.002$ & $6.476\pm0.085$ & $(1.130\pm0.030)\times10^{-1}$ & $23/12$\\
                              & X    & $1.100\pm0.023$ & $0.100\pm0.002$ & $6.476\pm0.129$ & $(4.500\pm0.010)\times10^{-2}$ & $29/12$\\
\hline
                              & I    & $0.255\pm0.004$ & $0.396\pm0.003$ & $10.007\pm0.060$ & $(9.700\pm0.069)\times10^{-2}$ & $7/9$\\
                              & II   & $0.255\pm0.003$ & $0.372\pm0.001$ & $9.615\pm0.043$  & $(7.000\pm0.060)\times10^{-2}$ & $10/9$\\
                              & III  & $0.255\pm0.006$ & $0.328\pm0.001$ & $8.700\pm0.026$  & $(6.000\pm0.060)\times10^{-2}$ & $4/9$\\
$\Xi^{-}+\overline\Xi{}^{+}$  & IV   & $0.255\pm0.005$ & $0.320\pm0.001$ & $8.553\pm0.034$  & $(4.800\pm0.019)\times10^{-2}$ & $2/9$\\
$(ssd,\bar{s}\bar{s}\bar{d})$ & V    & $0.255\pm0.005$ & $0.296\pm0.001$ & $7.629\pm0.027$  & $(3.999\pm0.010)\times10^{-2}$ & $4/9$\\
$(1/N_{evt}) d^2N/dp_{T}dy$   & VI   & $0.255\pm0.010$ & $0.276\pm0.001$ & $7.366\pm0.035$  & $(3.400\pm0.020)\times10^{-2}$ & $3/9$\\
$[({\rm GeV}/c)^{-1}]$        & VII  & $0.255\pm0.005$ & $0.261\pm0.001$ & $7.306\pm0.031$  & $(2.579\pm0.025)\times10^{-2}$ & $4/9$\\
                              & VIII & $0.255\pm0.007$ & $0.242\pm0.001$ & $6.844\pm0.035$  & $(1.899\pm0.010)\times10^{-2}$ & $4/9$\\
                              & IX   & $0.255\pm0.006$ & $0.210\pm0.001$ & $6.255\pm0.018$  & $(1.240\pm0.012)\times10^{-2}$ & $2/9$\\
                              & X    & $0.255\pm0.009$ & $0.175\pm0.001$ & $6.060\pm0.033$  & $(5.399\pm0.199)\times10^{-3}$ & $4/9$\\
\hline
$\Omega^{-}+\overline{\Omega}{}^{+}$ & I+II     & $0.830\pm0.004$ & $0.312\pm0.002$ & $9.597\pm0.031$ & $(8.099\pm0.010)\times10^{-3}$ & $1/2$\\
$(sss,\bar{s}\bar{s}\bar{s})$        & III+IV   & $0.830\pm0.004$ & $0.288\pm0.002$ & $9.548\pm0.033$ & $(4.899\pm0.010)\times10^{-3}$ & $2/2$\\
$(1/N_{evt}) d^2N/dp_{T}dy$          & V+VI     & $0.830\pm0.005$ & $0.270\pm0.002$ & $9.572\pm0.032$ & $(2.899\pm0.010)\times10^{-3}$ & $1/2$\\
$[({\rm GeV}/c)^{-1}]$               & VII+VIII & $0.830\pm0.004$ & $0.190\pm0.002$ & $6.477\pm0.030$ & $(1.820\pm0.009)\times10^{-3}$ & $1/2$\\
                                     & IX+X     & $0.830\pm0.004$ & $0.169\pm0.002$ & $6.477\pm0.031$ & $(5.200\pm0.001)\times10^{-4}$ & $3/2$\\
\hline
\end{tabular}%
\end{center}}
\end{table*}

\begin{table*}[htbp]\vspace{0.2cm}
\justifying\noindent {\small Table 2. Values of $a_{0}$, $T$, $n$,
$N_0$, $\chi^2$, and ndof corresponding to the curves in Figure 2
for $pp$ collisions at $\sqrt{s}=13$ TeV. The particle type (quark
structure), spectrum form, and multiplicity classes are given
together. The multiplicity classes I, II, III, IV, V, VI, VII,
VIII, IX, and X for $\pi^++\pi^-$, $K^++K^-$, and $p+\overline p$
correspond to the centrality classes 0--0.92\%, 0.92--4.6\%,
4.6--9.2\%, 9.2--13.8\%, 13.8--18.4\%, 18.4--27.6\%, 27.6--36.8\%,
36.8--46\%, 46--64.5\%, and 64.5--100\%, respectively, and for
$K_S^0$, $K^{*0}$, $\Lambda+\overline\Lambda$,
$\Xi^{-}+\overline\Xi{}^{+}$, and
$\Omega^{-}+\overline{\Omega}{}^{+}$ correspond to the centrality
classes 0--0.9\%, 0.9--4.5\%, 4.5--8.9\%, 8.9--13.5\%, 13.5--18\%,
18--27\%, 27--36.1\%, 36.1--45.3\%, 45.3--64.5\%, and 64.5--100\%,
respectively. \vspace{-0.35cm}

\begin{center}
\newcommand{\tabincell}[2]{\begin{tabular}{@{}#1@{}}#2\end{tabular}}
\begin{tabular} {cccccccccccc}\\ \hline\hline
\tabincell{c}{Particle (quark structure)\\ and spectrum form} & \tabincell{c}{Multiplicity\\ class} & $a_{0}$ & $T$ (GeV) & $n$ & $N_0$ & $\chi^2$/ndof \\
\hline
                              & I    & $-0.550\pm0.003$ & $0.375\pm0.002$ & $4.996\pm0.045$ & $(2.599\pm0.070)\times10^{1}$ & $56/47$\\
                              & II   & $-0.550\pm0.002$ & $0.350\pm0.002$ & $4.945\pm0.022$ & $(2.030\pm0.030)\times10^{1}$ & $74/47$\\
                              & III  & $-0.550\pm0.004$ & $0.332\pm0.004$ & $4.870\pm0.021$ & $(1.670\pm0.020)\times10^{1}$ & $70/47$\\
$\pi^{+}+\pi^{-}$             & IV   & $-0.550\pm0.005$ & $0.326\pm0.002$ & $4.874\pm0.026$ & $(1.400\pm0.019)\times10^{1}$ & $60/47$\\
$(u\bar{d},d\bar{u})$         & V    & $-0.550\pm0.005$ & $0.315\pm0.002$ & $4.874\pm0.035$ & $(1.250\pm0.002)\times10^{1}$ & $55/47$\\
$(1/N_{evt}) d^2N/dp_{T}dy$   & VI   & $-0.550\pm0.005$ & $0.305\pm0.001$ & $7.366\pm0.020$ & $(1.030\pm0.014)\times10^{1}$ & $44/47$\\
$[({\rm GeV}/c)^{-1}]$        & VII  & $-0.550\pm0.002$ & $0.288\pm0.001$ & $4.870\pm0.016$ & $(8.229\pm0.089)\times10^{0}$ & $49/47$\\
                              & VIII & $-0.550\pm0.004$ & $0.272\pm0.001$ & $4.870\pm0.029$ & $(6.730\pm0.070)\times10^{0}$ & $56/47$\\
                              & IX   & $-0.550\pm0.004$ & $0.249\pm0.001$ & $4.865\pm0.028$ & $(4.840\pm0.050)\times10^{0}$ & $92/47$\\
                              & X    & $-0.550\pm0.005$ & $0.203\pm0.002$ & $4.861\pm0.022$ & $(2.800\pm0.050)\times10^{0}$ & $324/47$\\
\hline
                              & I    & $0.148\pm0.003$ & $0.287\pm0.003$ & $5.276\pm0.082$ & $(3.179\pm0.040)\times10^{0}$  & $15/42$\\
                              & II   & $0.148\pm0.002$ & $0.270\pm0.002$ & $5.273\pm0.021$ & $(2.580\pm0.040)\times10^{0}$  & $9/42$\\
                              & III  & $0.148\pm0.003$ & $0.262\pm0.001$ & $5.273\pm0.018$ & $(2.060\pm0.030)\times10^{0}$  & $14/42$\\
$K^{+}+K^{-}$                 & IV   & $0.148\pm0.005$ & $0.254\pm0.002$ & $5.250\pm0.021$ & $(1.740\pm0.019)\times10^{0}$  & $25/42$\\
$(u\bar{s},s\bar{u})$         & V    & $0.148\pm0.005$ & $0.243\pm0.001$ & $5.180\pm0.022$ & $(1.540\pm0.020)\times10^{0}$  & $24/42$\\
$(1/N_{evt}) d^2N/dp_{T}dy$   & VI   & $0.148\pm0.005$ & $0.237\pm0.003$ & $5.242\pm0.022$ & $(1.250\pm0.019)\times10^{0}$  & $43/42$\\
$[({\rm GeV}/c)^{-1}]$        & VII  & $0.148\pm0.004$ & $0.224\pm0.001$ & $5.217\pm0.023$ & $(9.799\pm0.100)\times10^{-1}$ & $47/42$\\
                              & VIII & $0.148\pm0.003$ & $0.213\pm0.002$ & $5.223\pm0.025$ & $(7.700\pm0.080)\times10^{-1}$ & $54/42$\\
                              & IX   & $0.148\pm0.003$ & $0.196\pm0.002$ & $5.199\pm0.027$ & $(5.370\pm0.070)\times10^{-1}$ & $67/42$\\
                              & X    & $0.148\pm0.005$ & $0.158\pm0.002$ & $5.138\pm0.024$ & $(2.970\pm0.040)\times10^{-1}$ & $464/41$\\
\hline
                              & I    & $0.530\pm0.011$ & $0.224\pm0.002$ & $6.291\pm0.012$ & $(1.379\pm0.029)\times10^{0}$  & $24/40$\\
                              & II   & $0.530\pm0.013$ & $0.208\pm0.002$ & $6.278\pm0.096$ & $(1.060\pm0.020)\times10^{0}$  & $24/40$\\
                              & III  & $0.530\pm0.012$ & $0.197\pm0.002$ & $6.267\pm0.044$ & $(8.799\pm0.149)\times10^{-1}$ & $22/40$\\
$p+\overline{p}$              & IV   & $0.530\pm0.011$ & $0.193\pm0.001$ & $6.278\pm0.012$ & $(7.299\pm0.090)\times10^{-1}$ & $26/40$\\
$(uud,\bar{u}\bar{u}\bar{d})$ & V    & $0.530\pm0.013$ & $0.184\pm0.003$ & $6.262\pm0.033$ & $(6.499\pm0.079)\times10^{-1}$ & $20/40$\\
$(1/N_{evt}) d^2N/dp_{T}dy$   & VI   & $0.530\pm0.023$ & $0.175\pm0.001$ & $6.247\pm0.055$ & $(5.600\pm0.080)\times10^{-1}$ & $27/40$\\
$[({\rm GeV}/c)^{-1}]$        & VII  & $0.530\pm0.022$ & $0.165\pm0.002$ & $6.240\pm0.031$ & $(4.300\pm0.050)\times10^{-1}$ & $31/40$\\
                              & VIII & $0.530\pm0.015$ & $0.155\pm0.002$ & $6.228\pm0.033$ & $(3.450\pm0.060)\times10^{-1}$ & $40/40$\\
                              & IX   & $0.530\pm0.009$ & $0.141\pm0.001$ & $6.156\pm0.036$ & $(2.360\pm0.040)\times10^{-1}$ & $48/40$\\
                              & X    & $0.530\pm0.005$ & $0.109\pm0.001$ & $5.914\pm0.028$ & $(1.249\pm0.029)\times10^{-1}$ & $36/40$\\
\hline
                              & I    & $0.148\pm0.003$ & $0.304\pm0.002$ & $5.607\pm0.033$ & $(1.610\pm0.019)\times10^{0}$  & $18/34$\\
                              & II   & $0.148\pm0.005$ & $0.288\pm0.002$ & $5.515\pm0.038$ & $(1.210\pm0.080)\times10^{0}$  & $33/34$\\
                              & III  & $0.148\pm0.004$ & $0.273\pm0.002$ & $5.461\pm0.063$ & $(9.899\pm0.079)\times10^{-1}$ & $37/34$\\
$K^{0}_{S}$                   & IV   & $0.148\pm0.003$ & $0.257\pm0.001$ & $5.332\pm0.032$ & $(8.500\pm0.040)\times10^{-1}$ & $27/34$\\
$(d\bar{s})$                  & V    & $0.148\pm0.005$ & $0.251\pm0.002$ & $5.302\pm0.024$ & $(7.500\pm0.080)\times10^{-1}$ & $35/34$\\
$(1/N_{evt}) d^2N/dp_{T}dy$   & VI   & $0.148\pm0.003$ & $0.243\pm0.001$ & $5.303\pm0.036$ & $(6.239\pm0.069)\times10^{-1}$ & $54/34$\\
$[({\rm GeV}/c)^{-1}]$        & VII  & $0.148\pm0.004$ & $0.234\pm0.002$ & $5.337\pm0.048$ & $(4.850\pm0.040)\times10^{-1}$ & $65/34$\\
                              & VIII & $0.148\pm0.004$ & $0.218\pm0.001$ & $5.285\pm0.043$ & $(3.930\pm0.050)\times10^{-1}$ & $64/34$\\
                              & IX   & $0.148\pm0.003$ & $0.200\pm0.001$ & $5.229\pm0.008$ & $(2.760\pm0.030)\times10^{-1}$ & $60/34$\\
                              & X    & $0.148\pm0.003$ & $0.158\pm0.001$ & $5.000\pm0.037$ & $(1.440\pm0.029)\times10^{-1}$ & $51/34$\\
\hline
\end{tabular}%
\end{center}}
\end{table*}

\begin{table*}[htb!]\vspace{0.2cm}
\justifying\noindent {\small Table 2. Continued. \vspace{-0.35cm}

\begin{center}
\newcommand{\tabincell}[2]{\begin{tabular}{@{}#1@{}}#2\end{tabular}}
\begin{tabular} {cccccccccccc}\\ \hline\hline
\tabincell{c}{Particle (quark structure)\\ and spectrum form}& \tabincell{c}{Multiplicity\\ class} & $a_{0}$ & $T$ (GeV) & $n$ & $N_0$ & $\chi^2$/ndof \\
\hline
                              & I    & $1.450\pm0.022$ & $0.209\pm0.002$ & $6.978\pm0.013$ & $(4.569\pm0.150)\times10^{-1}$ & $2/8$\\
                              & II   & $1.450\pm0.028$ & $0.196\pm0.002$ & $6.925\pm0.008$ & $(3.619\pm0.080)\times10^{-1}$ & $8/10$\\
$K^{*0}$                      & III  & $1.450\pm0.035$ & $0.190\pm0.003$ & $6.805\pm0.013$ & $(2.899\pm0.070)\times10^{-1}$ & $8/10$\\
$(d\bar{s})$                  & IV+V & $1.450\pm0.030$ & $0.166\pm0.002$ & $6.318\pm0.016$ & $(2.379\pm0.010)\times10^{-1}$ & $16/10$\\
$(1/N_{evt}) d^2N/dp_{T}dy$   & VI   & $1.450\pm0.028$ & $0.166\pm0.004$ & $6.587\pm0.006$ & $(1.829\pm0.089)\times10^{-1}$ & $12/10$\\
$[({\rm GeV}/c)^{-1}]$        & VII  & $1.450\pm0.018$ & $0.153\pm0.005$ & $6.410\pm0.008$ & $(1.519\pm0.040)\times10^{-1}$ & $18/10$\\
                              & VIII & $1.450\pm0.025$ & $0.140\pm0.003$ & $6.415\pm0.013$ & $(1.219\pm0.050)\times10^{-1}$ & $17/10$\\
                              & IX   & $1.450\pm0.026$ & $0.127\pm0.002$ & $6.224\pm0.008$ & $(8.600\pm0.026)\times10^{-2}$ & $17/10$\\
                              & X    & $1.450\pm0.018$ & $0.103\pm0.003$ & $6.218\pm0.007$ & $(4.700\pm0.029)\times10^{-2}$ & $24/10$\\
\hline
                              & I    & $0.300\pm0.021$ & $0.357\pm0.003$ & $8.885\pm0.013$ & $(9.499\pm0.130)\times10^{-1}$ & $11/12$\\
                              & II   & $0.300\pm0.012$ & $0.328\pm0.004$ & $8.478\pm0.112$ & $(7.419\pm0.180)\times10^{-1}$ & $15/12$\\
                              & III  & $0.300\pm0.014$ & $0.306\pm0.002$ & $8.301\pm0.091$ & $(5.999\pm0.100)\times10^{-1}$ & $7/12$\\
$\Lambda+\overline{\Lambda}$  & IV   & $0.300\pm0.008$ & $0.287\pm0.003$ & $7.782\pm0.093$ & $(5.099\pm0.150)\times10^{-1}$ & $7/12$\\
$(uds,\bar{u}\bar{d}\bar{s})$ & V    & $0.300\pm0.011$ & $0.270\pm0.002$ & $7.355\pm0.043$ & $(4.500\pm0.050)\times10^{-1}$ & $5/12$\\
$(1/N_{evt}) d^2N/dp_{T}dy$   & VI   & $0.300\pm0.022$ & $0.257\pm0.001$ & $7.351\pm0.052$ & $(3.669\pm0.080)\times10^{-1}$ & $10/12$\\
$[({\rm GeV}/c)^{-1}]$        & VII  & $0.300\pm0.019$ & $0.223\pm0.002$ & $6.553\pm0.063$ & $(3.060\pm0.049)\times10^{-1}$ & $8/12$\\
                              & VIII & $0.300\pm0.014$ & $0.213\pm0.003$ & $6.554\pm0.081$ & $(2.339\pm0.040)\times10^{-1}$ & $7/12$\\
                              & IX   & $0.300\pm0.012$ & $0.192\pm0.001$ & $6.456\pm0.043$ & $(1.560\pm0.040)\times10^{-1}$ & $10/12$\\
                              & X    & $0.300\pm0.025$ & $0.152\pm0.001$ & $6.000\pm0.019$ & $(6.899\pm0.026)\times10^{-2}$ & $15/12$\\
\hline
                              & I    & $0.255\pm0.008$ & $0.431\pm0.002$ & $10.059\pm0.048$ & $(1.299\pm0.013)\times10^{-1}$ & $6/9$\\
                              & II   & $0.255\pm0.007$ & $0.405\pm0.003$ & $10.013\pm0.093$ & $(9.600\pm0.140)\times10^{-2}$ & $5/9$\\
                              & III  & $0.255\pm0.013$ & $0.381\pm0.003$ & $9.603\pm0.190$  & $(7.599\pm0.219)\times10^{-2}$ & $9/9$\\
$\Xi^{-}+\overline\Xi{}^{+}$  & IV   & $0.255\pm0.011$ & $0.371\pm0.002$ & $9.590\pm0.112$  & $(6.099\pm0.120)\times10^{-2}$ & $6/9$\\
$(ssd,\bar{s}\bar{s}\bar{d})$ & V    & $0.255\pm0.021$ & $0.353\pm0.003$ & $9.558\pm0.170$  & $(5.399\pm0.140)\times10^{-2}$ & $14/9$\\
$(1/N_{evt}) d^2N/dp_{T}dy$   & VI   & $0.255\pm0.012$ & $0.339\pm0.002$ & $9.557\pm0.087$  & $(4.299\pm0.130)\times10^{-2}$ & $14/9$\\
$[({\rm GeV}/c)^{-1}]$        & VII  & $0.255\pm0.013$ & $0.317\pm0.002$ & $8.985\pm0.152$  & $(3.199\pm0.080)\times10^{-2}$ & $12/9$\\
                              & VIII & $0.255\pm0.007$ & $0.280\pm0.002$ & $8.000\pm0.122$  & $(2.600\pm0.070)\times10^{-2}$ & $13/9$\\
                              & IX   & $0.255\pm0.004$ & $0.242\pm0.001$ & $6.593\pm0.088$  & $(1.600\pm0.010)\times10^{-2}$ & $4/9$\\
                              & X    & $0.255\pm0.012$ & $0.201\pm0.002$ & $6.587\pm0.080$  & $(5.900\pm0.160)\times10^{-3}$ & $7/9$\\
\hline
$\Omega^{-}+\overline{\Omega}{}^{+}$ & I+II     & $0.830\pm0.021$ & $0.335\pm0.001$ & $9.656\pm0.013$ & $(1.020\pm0.003)\times10^{-2}$  & $1/2$\\
$(sss,\bar{s}\bar{s}\bar{s})$        & III+IV   & $0.830\pm0.018$ & $0.290\pm0.001$ & $9.640\pm0.013$ & $(4.100\pm0.200)\times10^{-3}$  & $5/2$\\
$(1/N_{evt}) d^2N/dp_{T}dy$          & V+VI     & $0.830\pm0.020$ & $0.288\pm0.001$ & $9.640\pm0.016$ & $(4.099\pm0.150)\times10^{-3}$  & $4/2$\\
$[({\rm GeV}/c)^{-1}]$               & VII+VIII & $0.830\pm0.004$ & $0.276\pm0.002$ & $9.640\pm0.008$ & $(1.9000\pm0.080)\times10^{-3}$ & $8/2$\\\
                                     & IX+X     & $0.830\pm0.022$ & $0.233\pm0.003$ & $9.622\pm0.008$ & $(6.299\pm0.150)\times10^{-4}$  & $1/1$\\
\hline
\end{tabular}%
\end{center}}
\end{table*}

\begin{table*}[htbp]\vspace{-.5cm}
\justifying\noindent {\small Table 3. Values of $a_{0}$, $T$, $n$,
$N_0$, $\chi^2$, and ndof corresponding to the curves in Figure 3
for p--Pb collisions at $\sqrt{s_{\rm NN}}=5.02$ TeV. The particle
type (quark structure), spectrum form, and centrality are given
together. \vspace{-0.35cm}

\begin{center}
\newcommand{\tabincell}[2]{\begin{tabular}{@{}#1@{}}#2\end{tabular}}
\begin{tabular} {cccccccccccc}\\ \hline\hline
\tabincell{c}{Particle (quark structure)\\ spectrum form}& Centrality & $a^{0}$ & $T$ (GeV) & $n$ & $N_0$ & $\chi^2$/ndof \\
\hline
                                       & 0--5\%    & $-0.550\pm0.003$ & $0.382\pm0.003$ & $6.244\pm0.063$ & $(2.130\pm0.025)\times10^{1}$ & $148/54$\\
$\pi^{+}+\pi^{-}$                      & 5--10\%   & $-0.550\pm0.004$ & $0.370\pm0.002$ & $6.004\pm0.072$ & $(1.743\pm0.028)\times10^{1}$ & $135/54$\\
$(u\bar{d},d\bar{u})$                  & 10--20\%  & $-0.550\pm0.004$ & $0.358\pm0.001$ & $5.619\pm0.121$ & $(1.461\pm0.025)\times10^{1}$ & $148/54$\\
$(1/N_{evt}2 \pi p_{T}) d^2N/dp_{T}dy$ & 20--40\%  & $-0.550\pm0.005$ & $0.344\pm0.001$ & $5.616\pm0.030$ & $(1.153\pm0.019)\times10^{1}$ & $142/54$\\
$[({\rm GeV}/c)^{-2}]$                 & 40--60\%  & $-0.550\pm0.004$ & $0.317\pm0.002$ & $5.330\pm0.034$ & $(8.168\pm0.251)\times10^{0}$ & $130/54$\\
                                       & 60--80\%  & $-0.550\pm0.002$ & $0.300\pm0.002$ & $5.332\pm0.033$ & $(5.027\pm0.094)\times10^{0}$ & $93/54$\\
                                       & 80--100\% & $-0.550\pm0.003$ & $0.255\pm0.001$ & $5.032\pm0.038$ & $(2.199\pm0.072)\times10^{0}$ & $173/54$\\
\hline
                                       & 0--5\%    & $0.220\pm0.007$ & $0.265\pm0.002$ & $5.968\pm0.021$ & $(2.859\pm0.060)\times10^{0}$  & $63/47$\\
$K^{+}+K^{-}$                          & 5--10\%   & $0.220\pm0.013$ & $0.260\pm0.004$ & $5.948\pm0.083$ & $(2.356\pm0.063)\times10^{0}$  & $36/47$\\
$(u\bar{s},s\bar{u})$                  & 10--20\%  & $0.220\pm0.006$ & $0.254\pm0.002$ & $5.872\pm0.061$ & $(1.948\pm0.038)\times10^{0}$  & $24/47$\\
$(1/N_{evt}2 \pi p_{T}) d^2N/dp_{T}dy$ & 20--40\%  & $0.220\pm0.006$ & $0.253\pm0.002$ & $5.878\pm0.046$ & $(1.477\pm0.025)\times10^{0}$  & $11/47$\\
$[({\rm GeV}/c)^{-2}]$                 & 40--60\%  & $0.220\pm0.003$ & $0.241\pm0.001$ & $5.741\pm0.053$ & $(9.990\pm0.031)\times10^{-1}$ & $28/47$\\
                                       & 60--80\%  & $0.220\pm0.006$ & $0.216\pm0.003$ & $5.446\pm0.048$ & $(5.969\pm0.157)\times10^{-1}$ & $32/47$\\
                                       & 80--100\% & $0.220\pm0.007$ & $0.190\pm0.002$ & $5.446\pm0.056$ & $(2.419\pm0.062)\times10^{-1}$ & $88/47$\\
\hline
                                       & 0--5\%   & $0.398\pm0.021$ & $0.283\pm0.003$ & $8.270\pm0.063$ & $(2.262\pm0.075)\times10^{0}$  & $56/45$\\
$p+\overline{p}$                       & 5--10\%  & $0.398\pm0.012$ & $0.274\pm0.002$ & $8.155\pm0.120$ & $(1.847\pm0.062)\times10^{0}$  & $34/45$\\
$(uud,\bar{u}\bar{u}\bar{d})$          & 10--20\% & $0.398\pm0.024$ & $0.267\pm0.004$ & $7.900\pm0.111$ & $(1.558\pm0.037)\times10^{0}$  & $33/45$\\
$(1/N_{evt}2 \pi p_{T}) d^2N/dp_{T}dy$ & 20--40\% & $0.398\pm0.021$ & $0.256\pm0.004$ & $7.885\pm0.112$ & $(1.206\pm0.019)\times10^{0}$  & $18/45$\\
$[({\rm GeV}/c)^{-2}]$                 & 40--60\% & $0.398\pm0.024$ & $0.236\pm0.004$ & $7.652\pm0.052$ & $(8.357\pm0.125)\times10^{-1}$ & $36/45$\\
                                       & 60--80\% & $0.398\pm0.029$ & $0.208\pm0.003$ & $7.145\pm0.066$ & $(5.152\pm0.144)\times10^{-1}$ & $45/45$\\
\hline
                                       & 0--5\%    & $0.220\pm0.005$ & $0.279\pm0.002$ & $6.451\pm0.052$ & $(1.445\pm0.028)\times10^{0}$  & $13/30$\\
$K^{0}_{S}$                            & 5--10\%   & $0.220\pm0.007$ & $0.279\pm0.003$ & $6.436\pm0.068$ & $(1.162\pm0.028)\times10^{0}$  & $14/30$\\
$(d\bar{s})$                           & 10--20\%  & $0.220\pm0.010$ & $0.278\pm0.002$ & $6.426\pm0.049$ & $(9.581\pm0.157)\times10^{-1}$ & $18/30$\\
$(1/N_{evt}2 \pi p_{T}) d^2N/dp_{T}dy$ & 20--40\%  & $0.220\pm0.008$ & $0.260\pm0.002$ & $6.000\pm0.062$ & $(7.383\pm0.219)\times10^{-1}$ & $20/30$\\
$[({\rm GeV}/c)^{-2}]$                 & 40--60\%  & $0.220\pm0.010$ & $0.244\pm0.002$ & $5.762\pm0.061$ & $(5.027\pm0.126)\times10^{-1}$ & $21/30$\\
                                       & 60--80\%  & $0.220\pm0.013$ & $0.217\pm0.001$ & $5.527\pm0.058$ & $(3.079\pm0.072)\times10^{-1}$ & $45/30$\\
                                       & 80--100\% & $0.220\pm0.011$ & $0.181\pm0.002$ & $5.125\pm0.059$ & $(1.319\pm0.053)\times10^{-1}$ & $33/30$\\
\hline
$\Sigma(1385)^+$$(suu)$                & 0--20\%   & $1.300\pm0.006$ & $0.201\pm0.002$ & $7.682\pm0.065$ & $(4.441\pm0.050)\times10^{-2}$ & $2/2$\\
$(1/N_{evt}2 \pi p_{T}) d^2N/dp_{T}dy$ & 20--60\%  & $1.300\pm0.004$ & $0.183\pm0.001$ & $7.661\pm0.056$ & $(2.396\pm0.035)\times10^{-2}$ & $4/2$\\
$[({\rm GeV}/c)^{-2}]$                 & 60--100\% & $1.300\pm0.008$ & $0.157\pm0.002$ & $7.523\pm0.077$ & $(6.596\pm0.035)\times10^{-3}$ & $9/2$\\
\hline
                                        & 0--5\%    & $1.300\pm0.010$ & $0.187\pm0.001$ & $8.512\pm0.058$ & $(8.167\pm0.220)\times10^{-1}$ & $3/16$\\
$\Lambda+\overline{\Lambda}$            & 5--10\%   & $1.300\pm0.010$ & $0.181\pm0.001$ & $8.187\pm0.055$ & $(6.439\pm0.094)\times10^{-1}$ & $4/16$\\
$(uds,\bar{u}\bar{d}\bar{s})$           & 10--20\%  & $1.300\pm0.011$ & $0.174\pm0.003$ & $7.949\pm0.063$ & $(5.403\pm0.063)\times10^{-1}$ & $2/16$\\
$(1/N_{evt}2 \pi p_{T}) d^2N/dp_{T}dy$  & 20--40\%  & $1.300\pm0.005$ & $0.165\pm0.001$ & $7.731\pm0.052$ & $(4.083\pm0.037)\times10^{-1}$ & $4/16$\\
$[({\rm GeV}/c)^{-2}]$                  & 40--60\%  & $1.300\pm0.018$ & $0.154\pm0.003$ & $7.631\pm0.089$ & $(2.764\pm0.069)\times10^{-1}$ & $12/16$\\
                                        & 60--80\%  & $1.300\pm0.023$ & $0.131\pm0.002$ & $6.958\pm0.083$ & $(1.633\pm0.041)\times10^{-1}$ & $10/16$\\
                                        & 80--100\% & $1.300\pm0.019$ & $0.111\pm0.002$ & $6.692\pm0.102$ & $(6.283\pm0.251)\times10^{-2}$ & $18/16$\\
\hline
                                        & 0--5\%    & $1.000\pm0.031$ & $0.259\pm0.003$ & $9.399\pm0.102$ & $(5.781\pm0.079)\times10^{-2}$ & $13/13$\\
$(\Xi^{-}+\overline\Xi^{+})/2$          & 5--10\%   & $1.000\pm0.019$ & $0.249\pm0.004$ & $8.964\pm0.103$ & $(4.618\pm0.110)\times10^{-2}$ & $9/13$\\
$(ssd,\bar{s}\bar{s}\bar{d})$           & 10--20\%  & $1.000\pm0.018$ & $0.244\pm0.003$ & $8.882\pm0.089$ & $(3.770\pm0.085)\times10^{-2}$ & $7/13$\\
$(1/N_{evt}2 \pi p_{T}) d^2N/dp_{T}dy$  & 20--40\%  & $1.000\pm0.007$ & $0.235\pm0.001$ & $8.824\pm0.086$ & $(2.733\pm0.050)\times10^{-2}$ & $6/13$\\
$[({\rm GeV}/c)^{-2}]$                  & 40--60\%  & $1.000\pm0.023$ & $0.222\pm0.002$ & $8.819\pm0.079$ & $(1.791\pm0.053)\times10^{-2}$ & $11/13$\\
                                        & 60--80\%  & $1.000\pm0.002$ & $0.203\pm0.001$ & $8.030\pm0.001$ & $(9.204\pm0.003)\times10^{-3}$ & $40/12$\\
                                        & 80--100\% & $1.000\pm0.000$ & $0.153\pm0.000$ & $6.324\pm0.001$ & $(3.299\pm0.003)\times10^{-3}$ & $11/11$\\
\hline
                                        & 0--5\%    & $0.830\pm0.008$ & $0.319\pm0.002$ & $8.346\pm0.062$ & $(6.597\pm0.087)\times10^{-3}$ & $4/4$\\
$(\Omega^{-}+\overline{\Omega}{}^{+})/2$& 5--10\%   & $0.830\pm0.006$ & $0.319\pm0.002$ & $8.139\pm0.088$ & $(5.655\pm0.094)\times10^{-3}$ & $3/4$\\
$(sss,\bar{s}\bar{s}\bar{s})$           & 10--20\%  & $0.830\pm0.021$ & $0.294\pm0.004$ & $7.722\pm0.103$ & $(4.398\pm0.069)\times10^{-3}$ & $7/4$\\
$(1/N_{evt}2 \pi p_{T}) d^2N/dp_{T}dy$  & 20--40\%  & $0.830\pm0.021$ & $0.278\pm0.002$ & $7.233\pm0.111$ & $(3.047\pm0.038)\times10^{-3}$ & $2/4$\\
$[({\rm GeV}/c)^{-2}]$                  & 40--60\%  & $0.830\pm0.008$ & $0.243\pm0.003$ & $6.832\pm0.193$ & $(1.885\pm0.063)\times10^{-3}$ & $7/4$\\
                                        & 60--80\%  & $0.830\pm0.011$ & $0.234\pm0.002$ & $6.901\pm0.098$ & $(1.005\pm0.019)\times10^{-2}$ & $5/4$\\
                                        & 80--100\% & $0.830\pm0.018$ & $0.182\pm0.003$ & $6.003\pm0.105$ & $(2.984\pm0.016)\times10^{-4}$ & $2/4$\\
\hline
\end{tabular}%
\end{center}}
\end{table*}

\begin{table*}[htbp]\vspace{-.5cm}
\vspace{0.25cm} \justifying\noindent {\small Table 4. Values of
$a^{0}$, $T$, $n$, $N_0$, $\chi^2$, and ndof corresponding to the
curves in Figure 4 for Pb--Pb collisions at $\sqrt{s_{\rm
NN}}=2.76$ TeV. The particle type (quark structure), spectrum
form, and centrality are given together. \vspace{-0.35cm}

\begin{center}
\newcommand{\tabincell}[2]{\begin{tabular}{@{}#1@{}}#2\end{tabular}}
\begin{tabular} {cccccccccccc}\\ \hline\hline
\tabincell{c}{Particle (quark structure)\\ spectrum form}& Centrality & $a_{0}$ & $T$ (GeV) & $n$ & $N_0$ & $\chi^2$/ndof \\
\hline
                                       & 0--5\%   & $-0.300\pm0.004$ & $0.236\pm0.001$ & $6.599\pm0.012$ & $(2.325\pm0.042)\times10^{3}$ & $562/59$\\
$\pi^{+}+\pi^{-}$                      & 5--10\%  & $-0.300\pm0.002$ & $0.236\pm0.002$ & $6.506\pm0.014$ & $(1.930\pm0.021)\times10^{3}$ & $469/59$\\
$(u\bar{d},d\bar{u})$                  & 10--20\% & $-0.300\pm0.004$ & $0.236\pm0.002$ & $6.226\pm0.023$ & $(1.414\pm0.040)\times10^{3}$ & $471/59$\\
$(1/N_{evt}2 \pi p_{T}) d^2N/dp_{T}dy$ & 20--40\% & $-0.300\pm0.005$ & $0.237\pm0.002$ & $6.206\pm0.022$ & $(8.043\pm0.302)\times10^{2}$ & $331/59$\\
$[({\rm GeV}/c)^{-2}]$                 & 40--60\% & $-0.300\pm0.004$ & $0.225\pm0.002$ & $5.850\pm0.025$ & $(3.086\pm0.101)\times10^{2}$ & $158/59$\\
                                       & 60--80\% & $-0.300\pm0.009$ & $0.218\pm0.004$ & $5.668\pm0.026$ & $(8.193\pm0.367)\times10^{1}$ & $99/59$\\
\hline
                                       & 0--5\%   & $0.300\pm0.008$ & $0.222\pm0.002$ & $7.198\pm0.031$ & $(3.177\pm0.093)\times10^{2}$ & $607/54$\\
$K^{+}+K^{-}$                          & 5--10\%  & $0.300\pm0.009$ & $0.221\pm0.001$ & $7.178\pm0.025$ & $(2.724\pm0.070)\times10^{2}$ & $524/54$\\
$(u\bar{s},s\bar{u})$                  & 10--20\% & $0.300\pm0.014$ & $0.216\pm0.003$ & $6.926\pm0.028$ & $(2.161\pm0.058)\times10^{2}$ & $415/54$\\
$(1/N_{evt}2 \pi p_{T}) d^2N/dp_{T}dy$ & 20--40\% & $0.300\pm0.022$ & $0.211\pm0.003$ & $6.526\pm0.032$ & $(1.190\pm0.035)\times10^{2}$ & $291/54$\\
$[({\rm GeV}/c)^{-2}]$                 & 40--60\% & $0.300\pm0.012$ & $0.202\pm0.005$ & $6.157\pm0.040$ & $(4.363\pm0.151)\times10^{1}$ & $95/54$\\
                                       & 60--80\% & $0.300\pm0.009$ & $0.201\pm0.003$ & $6.102\pm0.026$ & $(1.116\pm0.035)\times10^{1}$ & $25/54$\\
\hline
                                       & 0--5\%   & $1.000\pm0.005$ & $0.209\pm0.003$ & $10.357\pm0.023$ & $(1.089\pm0.050)\times10^{2}$  & $490/45$\\
$p+\overline{p}$                       & 5--10\%  & $1.000\pm0.020$ & $0.207\pm0.002$ & $10.337\pm0.035$ & $(9.178\pm0.402)\times10^{1}$  & $444/45$\\
$(uud,\bar{u}\bar{u}\bar{d})$          & 10--20\% & $1.000\pm0.015$ & $0.205\pm0.002$ & $10.077\pm0.080$ & $(6.977\pm0.312)\times10^{1}$  & $374/45$\\
$(1/N_{evt}2 \pi p_{T}) d^2N/dp_{T}dy$ & 20--40\% & $1.000\pm0.019$ & $0.193\pm0.004$ & $9.391\pm0.100$  &  $(3.921\pm0.090)\times10^{1}$ & $259/45$\\
$[({\rm GeV}/c)^{-2}]$                 & 40--60\% & $1.000\pm0.003$ & $0.170\pm0.003$ & $8.183\pm0.082$  &  $(1.553\pm0.060)\times10^{1}$ & $104/45$\\
                                       & 60--80\% & $1.000\pm0.002$ & $0.155\pm0.002$ & $8.005\pm0.007$  &  $(4.524\pm0.090)\times10^{0}$ & $14/45$\\
\hline
                                 & 0--5\%   & $0.220\pm0.033$ & $0.251\pm0.005$ & $8.022\pm0.012$ & $(1.003\pm0.050)\times10^{2}$  & $545/29$\\
$K^{0}_{S}$                      & 5--10\%  & $0.220\pm0.022$ & $0.251\pm0.005$ & $8.018\pm0.017$ & $(8.660\pm0.340)\times10^{1}$  & $406/29$\\
$(d\bar{s})$                     & 10--20\% & $0.220\pm0.018$ & $0.251\pm0.003$ & $8.018\pm0.068$ & $(6.930\pm0.300)\times10^{1}$  & $317/29$\\
$(1/N_{evt}) d^2N/dp_{T}dy$      & 20--40\% & $0.220\pm0.012$ & $0.239\pm0.004$ & $7.238\pm0.053$ & $(3.889\pm0.179)\times10^{1}$  & $229/29$\\
$[({\rm GeV}/c)^{-2}]$           & 40--60\% & $0.220\pm0.013$ & $0.222\pm0.002$ & $6.490\pm0.049$ & $(1.469\pm0.056)\times10^{1}$  & $61/29$\\
                                 & 60--80\% & $0.220\pm0.008$ & $0.214\pm0.002$ & $6.083\pm0.040$ & $(3.579\pm0.039)\times10^{0}$  & $19/29$\\
                                 & 80--90\% & $0.220\pm0.009$ & $0.195\pm0.002$ & $5.719\pm0.067$ & $(7.959\pm0.219)\times10^{-1}$ & $22/28$\\
\hline
                                 & 0--5\%   & $1.950\pm0.009$ & $0.156\pm0.004$ & $8.468\pm0.064$ & $(1.728\pm0.094)\times10^{1}$ & $29/9$\\
$(K^{*0}+\overline{K}{}^{*0})/2$ & 5--10\%  & $1.950\pm0.031$ & $0.148\pm0.003$ & $8.276\pm0.061$ & $(1.521\pm0.125)\times10^{1}$ & $23/9$\\
$(u\bar{s},s\bar{u})$            & 10--20\% & $1.950\pm0.040$ & $0.148\pm0.003$ & $7.978\pm0.105$ & $(1.175\pm0.157)\times10^{1}$ & $50/9$\\
$(1/N_{evt}) d^2N/dp_{T}dy$      & 20--30\% & $1.950\pm0.026$ & $0.144\pm0.005$ & $7.646\pm0.203$ & $(8.734\pm0.817)\times10^{0}$ & $32/9$\\
$[({\rm GeV}/c)^{-1}]$           & 30--40\% & $1.950\pm0.019$ & $0.140\pm0.002$ & $7.646\pm0.104$ & $(6.723\pm0.565)\times10^{0}$ & $13/9$\\
                                 & 40--50\% & $1.950\pm0.018$ & $0.138\pm0.002$ & $7.613\pm0.066$ & $(4.398\pm0.163)\times10^{0}$ & $7/9$\\
\hline
                                 & 0--5\%   & $2.600\pm0.014$ & $0.125\pm0.002$ & $11.456\pm0.015$ &  $(2.403\pm0.060)\times10^{1}$  & $84/27$\\
$\Lambda$                        & 5--10\%  & $2.600\pm0.012$ & $0.124\pm0.001$ & $11.406\pm0.013$ &  $(2.170\pm0.060)\times10^{1}$  & $43/27$\\
$(uds)$                          & 10--20\% & $2.600\pm0.013$ & $0.127\pm0.002$ & $11.406\pm0.021$ &  $(1.740\pm0.090)\times10^{1}$  & $30/27$\\
$(1/N_{evt}) d^2N/dp_{T}dy$      & 20--40\% & $2.600\pm0.008$ & $0.110\pm0.001$ & $9.993\pm0.033$  &  $(9.899\pm0.299)\times10^{0}$  & $25/27$\\
$[({\rm GeV}/c)^{-1}]$           & 40--60\% & $2.600\pm0.006$ & $0.100\pm0.001$ & $9.367\pm0.045$  &  $(3.700\pm0.079)\times10^{0}$  & $5/27$\\
                                 & 60--80\% & $2.600\pm0.014$ & $0.085\pm0.001$ & $8.383\pm0.057$  &  $(8.800\pm0.099)\times10^{-1}$ & $21/27$\\
                                 & 80--90\% & $2.600\pm0.023$ & $0.078\pm0.001$ & $8.169\pm0.062$  &  $(1.720\pm0.069)\times10^{-1}$ & $37/25$\\
\hline
$(\Xi^{-}+\overline\Xi{}^{+})/2$ & 0--10\%  & $1.955\pm0.008$ & $0.194\pm0.002$ & $16.514\pm0.018$ & $(3.390\pm0.099)\times10^{0}$  & $68/23$\\
$(ssd,\bar{s}\bar{s}\bar{d})$    & 10--20\% & $1.955\pm0.010$ & $0.181\pm0.001$ & $14.343\pm0.089$ & $(2.579\pm0.059)\times10^{0}$  & $18/23$\\
$(1/N_{evt}) d^2N/dp_{T}dy$      & 20--40\% & $1.955\pm0.014$ & $0.168\pm0.001$ & $12.385\pm0.103$ & $(1.460\pm0.039)\times10^{0}$  & $21/23$\\
$[({\rm GeV}/c)^{-1}]$           & 40--60\% & $1.955\pm0.014$ & $0.164\pm0.002$ & $12.152\pm0.100$ & $(4.850\pm0.159)\times10^{-1}$ & $27/21$\\
                                 & 60--80\% & $1.955\pm0.025$ & $0.130\pm0.002$ & $9.000\pm0.013$  & $(1.099\pm0.019)\times10^{-1}$ & $18/21$\\
\hline
$(\Omega^{-}+\overline{\Omega}{}^{+})/2$ & 0--10\%  & $2.150\pm0.009$ & $0.186\pm0.002$ & $12.707\pm0.048$ & $(5.500\pm0.099)\times10^{-1}$ & $6/9$\\
$(sss,\bar{s}\bar{s}\bar{s})$            & 10--20\% & $2.150\pm0.008$ & $0.179\pm0.002$ & $12.492\pm0.056$ & $(4.099\pm0.109)\times10^{-1}$ & $8/9$\\
$(1/N_{evt}) d^2N/dp_{T}dy$              & 20--40\% & $2.150\pm0.008$ & $0.176\pm0.002$ & $12.400\pm0.088$ & $(2.280\pm0.140)\times10^{-1}$ & $10/9$\\
$[({\rm GeV}/c)^{-1}]$                   & 40--60\% & $2.150\pm0.011$ & $0.138\pm0.001$ & $9.008\pm0.066$  & $(7.599\pm0.129)\times10^{-2}$ & $3/8$\\
                                         & 60--80\% & $2.150\pm0.014$ & $0.136\pm0.002$ & $9.032\pm0.101$  & $(1.299\pm0.100)\times10^{-2}$ & $7/6$\\
\hline
\end{tabular}%
\end{center}}
\end{table*}

\section{Results and discussion}

\subsection{Comparison with data}

Figure 1 shows the multiplicity dependent $p_{T}$ spectra, the
double-differential yield $(1/N_{evt}) d^2N/dp_{T}dy$, of
$\pi^++\pi^-$ (a), $K^{+}+K^{-}$ (b), $p+\overline{p}$ (c),
$K^{0}_{S}$ (d), $K^{*}+\overline{K}{}^*$ (e),
$\Lambda+\overline{\Lambda}$ (f), $\Xi^{-}+\overline\Xi{}^{+}$
(g), and $\Omega^{-}+\overline{\Omega}{}^{+}$ (h) with rapidity
$|y|<0.5$, produced in $pp$ collisions at the center-of-mass
energy, $\sqrt{s}=7$ TeV, where $N_{evt}$ denotes the number of
events which can be omitted in the vertical axis according to the
format in the cited reference. Different symbols represent the
experimental data for different multiplicity classes determined by
the values of multiplicity (VOM) measured by the ALICE
Collaboration~\cite{44}, where in most cases the data are scaled
by constant multipliers marked in the panels for clarity. The
curves represent our fit results from the Monte Carlo
calculations. The values of free parameters ($a_0$, $T$, and $n$),
normalization constant ($N_0$), $\chi^{2}$, and number of degree
of freedom (ndof) are listed in Table 1 in which the particle
type, quark structure that makes up hadrons, and form of spectra
are mentioned. The corresponding serial numbers of VOM classes
marked with the Roman numerals in the figure are also listed in
Table 1 in terms of the percentage classes which can be regarded
as the centrality classes. One can see that our fit results are in
well agreement with the experiment data measured by the ALICE
Collaboration at midrapidity in $pp$ collisions at $\sqrt{s}=7$
TeV.

Similar to Figure 1, Figure 2 shows the multiplicity dependent
$p_{T}$ spectra, the double-differential yield $(1/N_{evt})
d^2N/dp_{T}dy$, of $\pi^++\pi^-$ (a), $K^{+}+K^{-}$ (b),
$p+\overline{p}$ (c), $K^{0}_{S}$ (d), $K^{*0}$ (e),
$\Lambda+\overline{\Lambda}$ (f), $\Xi^{-}+\overline\Xi{}^{+}$
(g), and $\Omega^{-}+\overline{\Omega}{}^{+}$ (h) with $|y|<0.5$,
produced in $pp$ collisions at $\sqrt{s}=13$ TeV measured by the
ALICE Collaboration~\cite{45,46,47}. Different symbols represent
the data for different serial numbers of VOM classes marked by the
Roman numerals, and the corresponding centrality classes are
listed in Table 2. The curves are our results of the Monte Carlo
calculations which are used to fit the data. Some of the data are
scaled by multiplying different amounts marked in the panels for
clarity. The values of $a_0$, $T$, $n$, $N$, $\chi^{2}$/ndof, and
other related information are listed in Table 2. From the figure
and $\chi^{2}$/ndof, one can see that our fit results are in good
agreement with the experimental data measured by the ALICE
Collaboration at midrapidity in $pp$ collisions at $\sqrt{s}=13$
TeV.

Similar to Figures 1 and 2, Figure 3 displays the centrality
dependent $p_{T}$ spectra, the invariant yield $(1/2\pi
p_{T})d^{2}N/dp_{T}dy$ (a--c) or $(1/N_{evt}2\pi
p_{T})d^{2}N/dp_{T}dy$ (d, f--h) or the double-differential yield
$(1/N_{evt})d^{2}N/dp_{T}dy$ (e), of $\pi^++\pi^-$ (a),
$K^{+}+K^{-}$ (b), $p+\overline{p}$ (c), $K^{0}_{S}$ (d),
$\Sigma(1385)^+$ (e), $\Lambda+\overline{\Lambda}$ (f),
$(\Xi^{-}+\overline\Xi{}^{+})/2$ (g), and
$(\Omega^{-}+\overline{\Omega}{}^{+})/2$ (h) with $-0.5<y<0$
(a--c, e, g, h) or $0<y<0.5$ (d, f), produced in $p$--Pb
collisions at $\sqrt{s_{NN}}=5.02$ TeV measured by the ALICE
Collaboration~\cite{10,48,49,50}. Different symbols represent the
experimental data with different centrality classes, where
different constant multipliers are used to re-scale the data for
clarity. The curves are our fit results based on the Monte Carlo
calculations. The values of $a_0$, $T$, $n$, $N_0$, and
$\chi^{2}$/ndof are listed in Table 3 with other information. One
can see that our fit results are in good agreement with the
experimental data measured by the ALICE Collaboration at mid-$y$
in $p$--Pb collisions at $\sqrt{s_{\rm NN}}=5.02$ TeV.

Similar to Figures 1--3, Figure 4 gives the centrality dependent
$p_T$ spectra, the invariant yield $(1/2\pi p_{T})d^{2}N/dp_{T}dy$
(a--c, e) or the double-differential yield
$(1/N_{evt})d^{2}N/dp_{T}dy$ (d, f--h), of $\pi^++\pi^-$ (a),
$K^{+}+K^{-}$ (b), $p+\overline{p}$ (c), $K^0_S$ (d),
$(K^{*0}+\overline{K}{}^{*0})/2$ (e), $\Lambda$ (f),
$(\Xi^{-}+\overline{\Xi}{}^{+})/2$ (g), and
$(\Omega^{-}+\overline{\Omega}{}^{+})/2$ (h) with $|\eta|<0.8$
(a--c) or $|y|<0.5$ (d--h), produced in Pb--Pb collisions at
$\sqrt{s_{NN}}=2.76$ TeV measured by the ALICE
Collaboration~\cite{51,52,53,54}. The symbols represent the
experimental data and the curves are our fit results. The value of
$a_0$, $T$, $n$, $N_0$, and $\chi^{2}$/ndof are listed in Table 4.
One can see that our fit results are approximately in agreement
with the experimental data measured by the ALICE Collaborations at
mid-$\eta$ or mid-$y$ in Pb--Pb collisions at $\sqrt{s_{\rm
NN}}=2.76$ TeV.

From the above comparisons one can see that the multi-source
thermal model at the quark level can fit the $p_T$ spectra of
identified hadrons produced at midrapidity in $pp$,
$p$--Pb, and Pb--Pb collisions at the LHC. We note that the degree
of fit for $pp$ collisions is better than that for $p$--Pb
collisions, and the degree of fit for $p$--Pb collisions is better
than that for Pb--Pb collisions. The phenomenon that the degree of
fit for small system is better than that for large system is
caused by the multiple scattering in large system. Due to the
multiple scattering, there are more factors affecting particle
production. In particular, in central Pb--Pb collisions, for light
particles ($\pi$, $K$, and $p$) with high-$p_T$, the departure of
the fit from data is more obvious. This is indeed caused by
multiple scattering and other medium effects.

In the model, the contribution of each quark to $p_T$ of given
hadron is assumed to obey the TP-like function with isotropic
azimuth. Although the analytical expression of $p_T$ distribution
is not available, the Monte Carlo method is performed in the
calculations. As we know, for a wide $p_T$ range, there are at
least two components namely the soft and hard components in the
structure of $p_T$ spectrum. The present work shows that we do not
need to distinguish the two components. Instead, we may use a set
of parameters to fit a wide $p_T$ spectrum, though the parameters
are multi-factor dependent.

\subsection{Tendencies of parameters}

\begin{figure*}[htbp]
\begin{center}
\includegraphics[width=11.0cm]{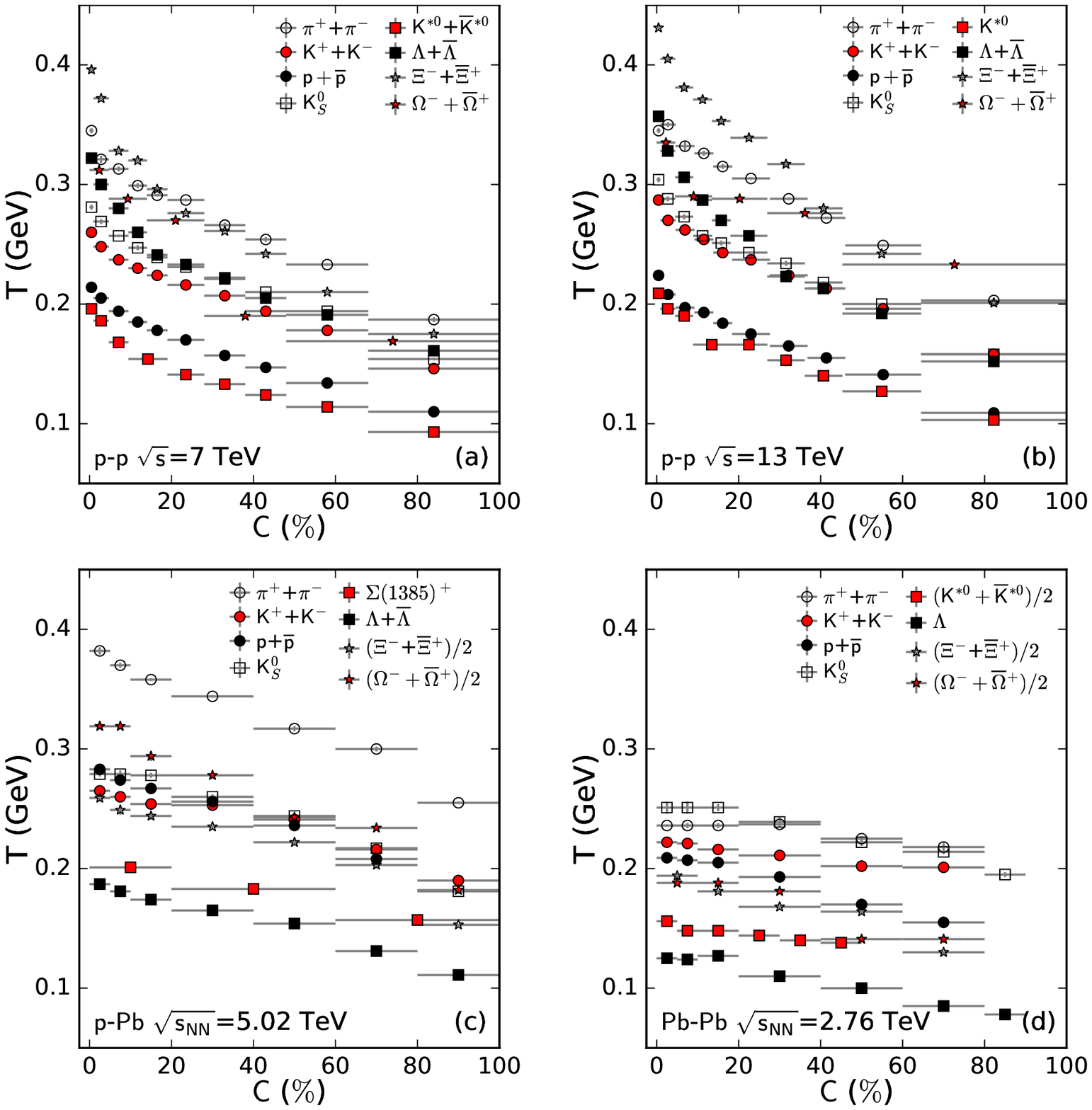}
\end{center}
\justifying\noindent {Fig. 5. Dependence of $T$ on $C$ for $pp$
collisions at $\sqrt{s}=7$ TeV (a), $pp$ collisions at $\sqrt{s}$
= 13 TeV (b), $p$--Pb collisions at $\sqrt{s_{\rm NN}}=5.02$ TeV
(c), and Pb--Pb collisions at $\sqrt{s_{\rm NN}}=2.76$ TeV (d).
Different symbols represent the parameters from the $p_T$ spectra
of different particles marked in the panels.}
\end{figure*}

\begin{figure*}[htbp]
\begin{center}
\includegraphics[width=11.0cm]{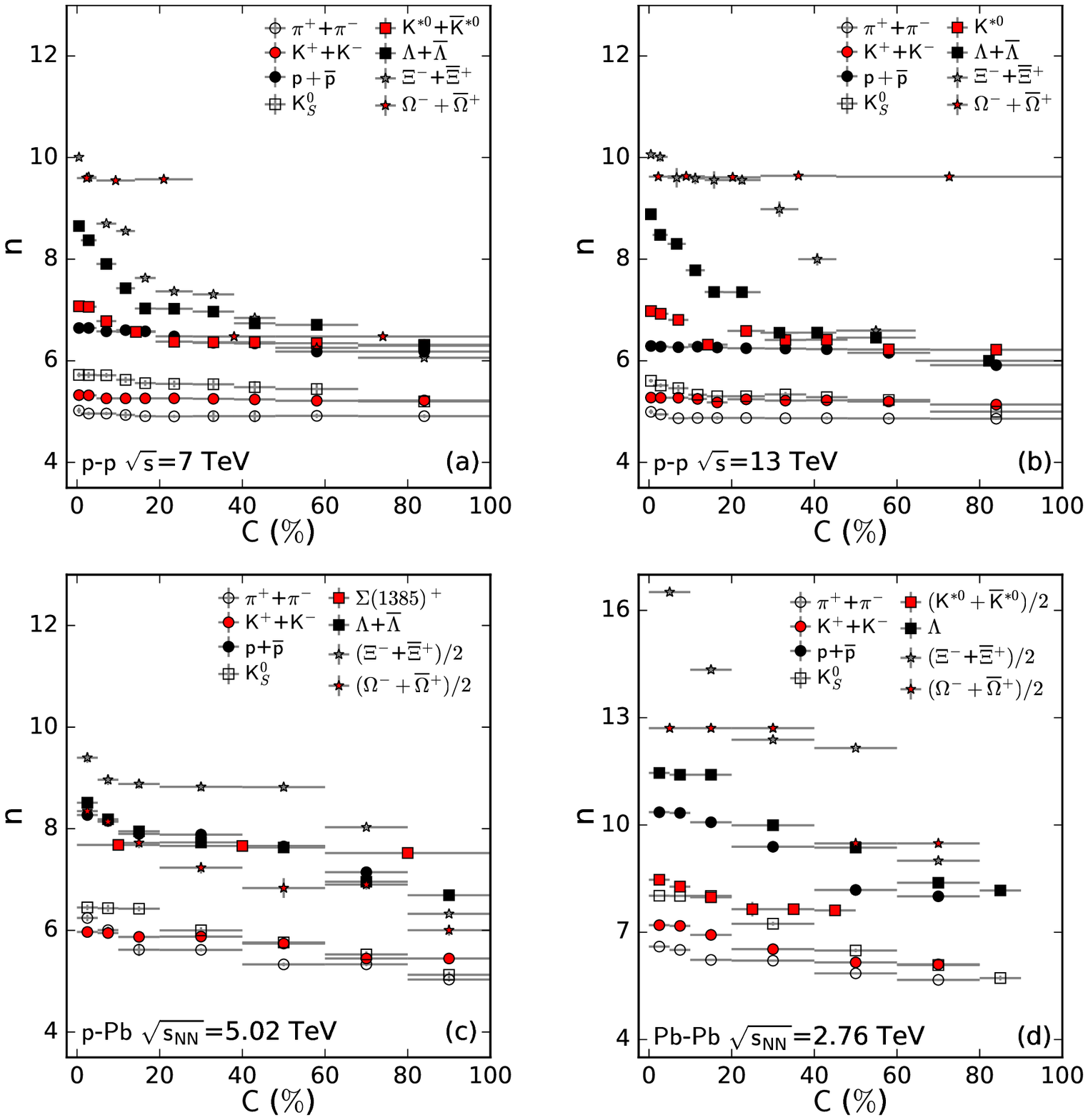}
\end{center}
\justifying\noindent {Fig. 6. Same as Figure 5, but showing the
dependence of $n$ on $C$.}
\end{figure*}

\begin{figure*}[htbp]
\begin{center}
\includegraphics[width=11.0cm]{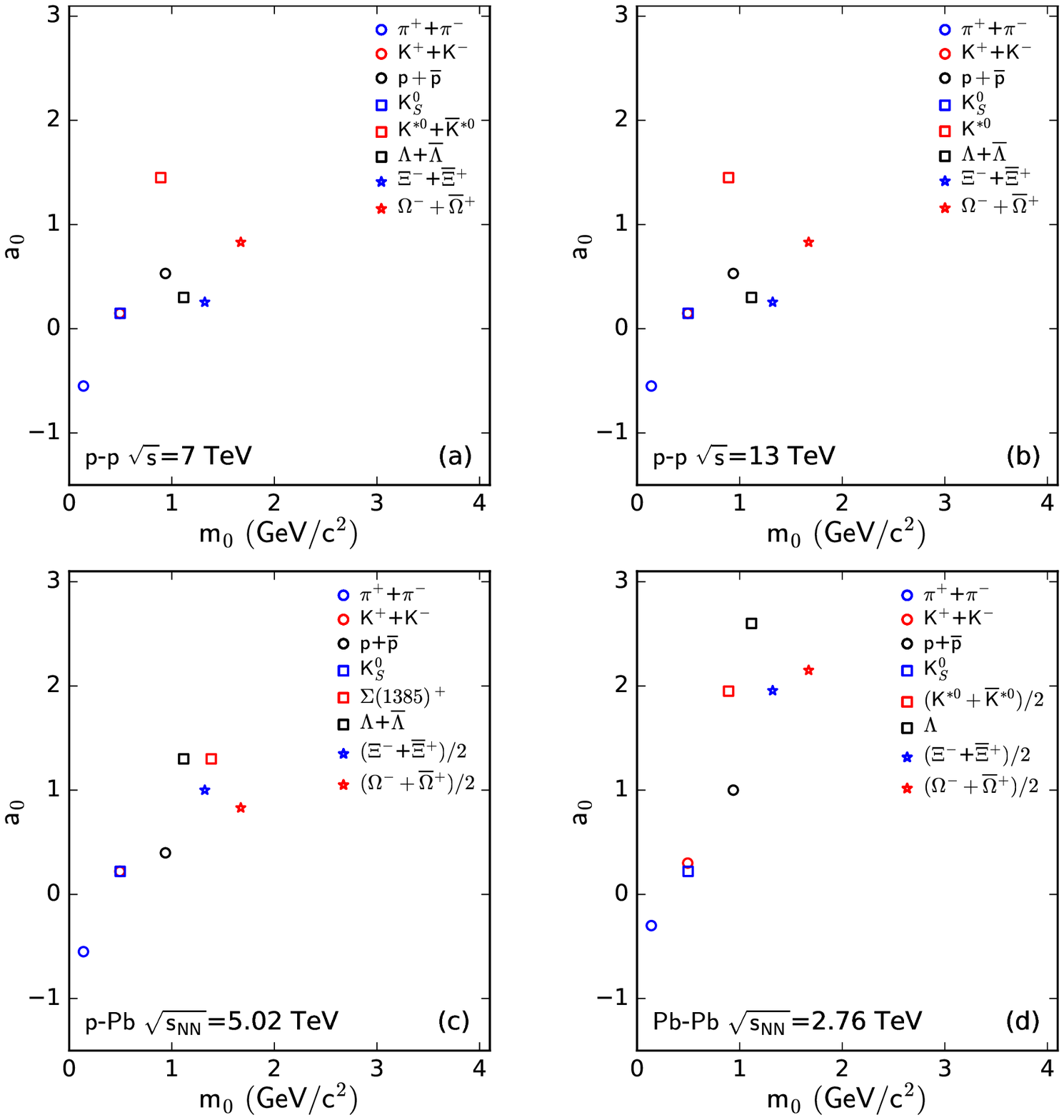}
\end{center}
\justifying\noindent {Fig. 7. Same as Figure 5, but showing the
dependence of $a_0$ on $m_0$.}
\end{figure*}

\begin{figure*}[htbp]
\begin{center}
\includegraphics[width=11.0cm]{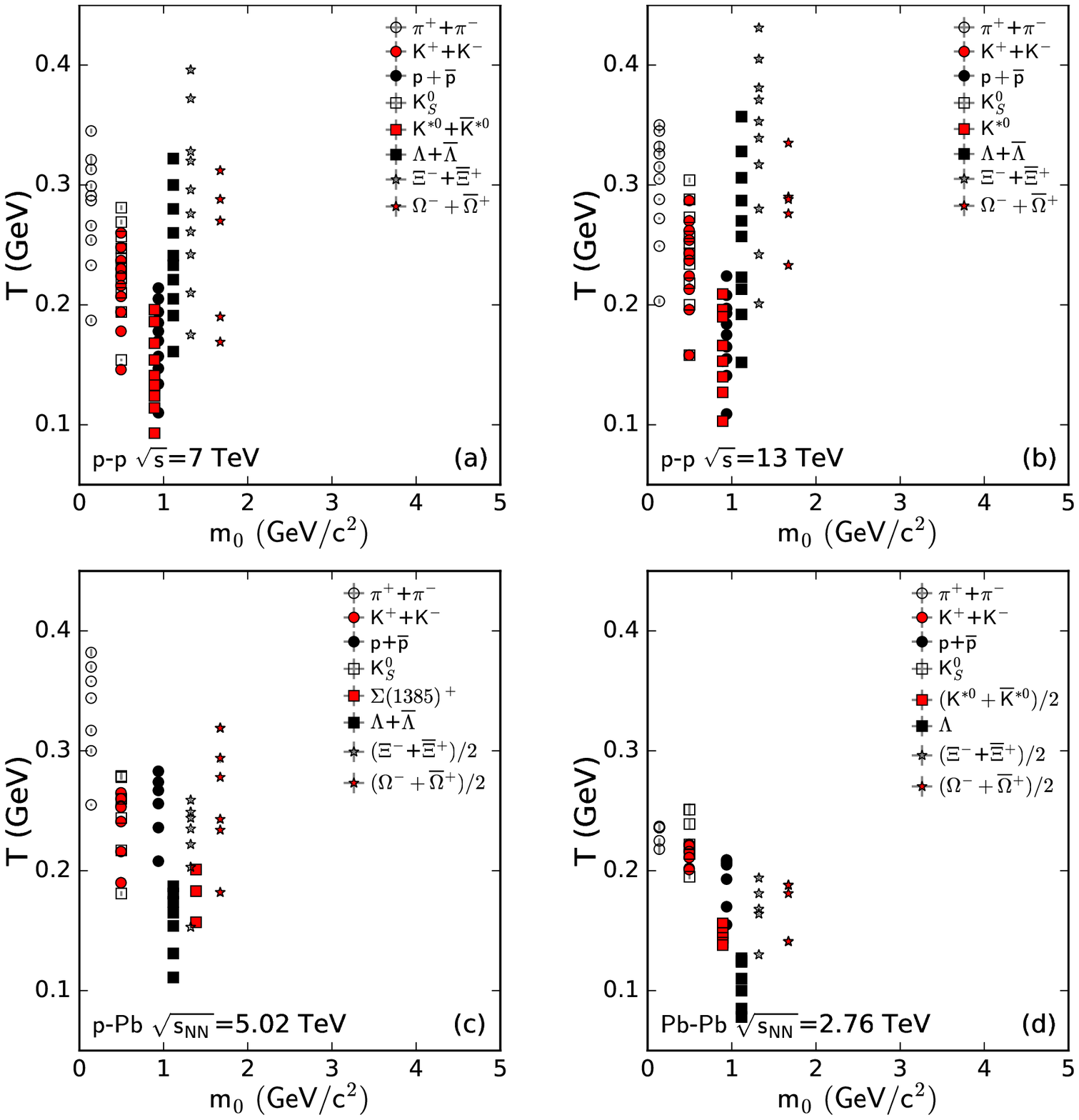}
\end{center}
\justifying\noindent {Fig. 8. Same as Figure 5, but showing the
dependence of $T$ on $m_0$.}
\end{figure*}

\begin{figure*}[htbp]
\begin{center}
\includegraphics[width=11.0cm]{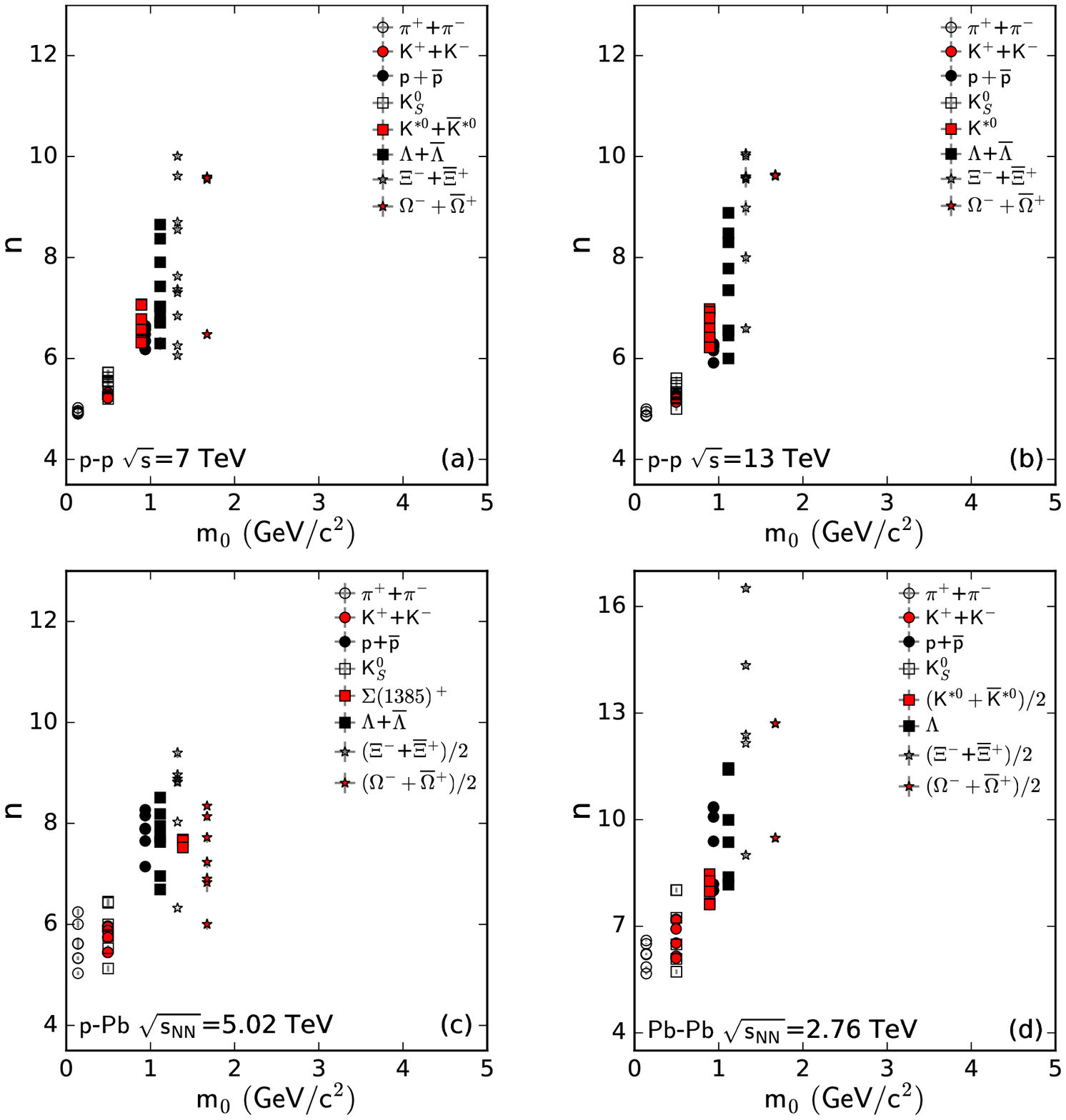}
\end{center}
\justifying\noindent {Fig. 9. Same as Figure 5, but showing the
dependence of $n$ on $m_0$.}
\end{figure*}

\begin{figure*}[htbp]
\begin{center}
\includegraphics[width=11.0cm]{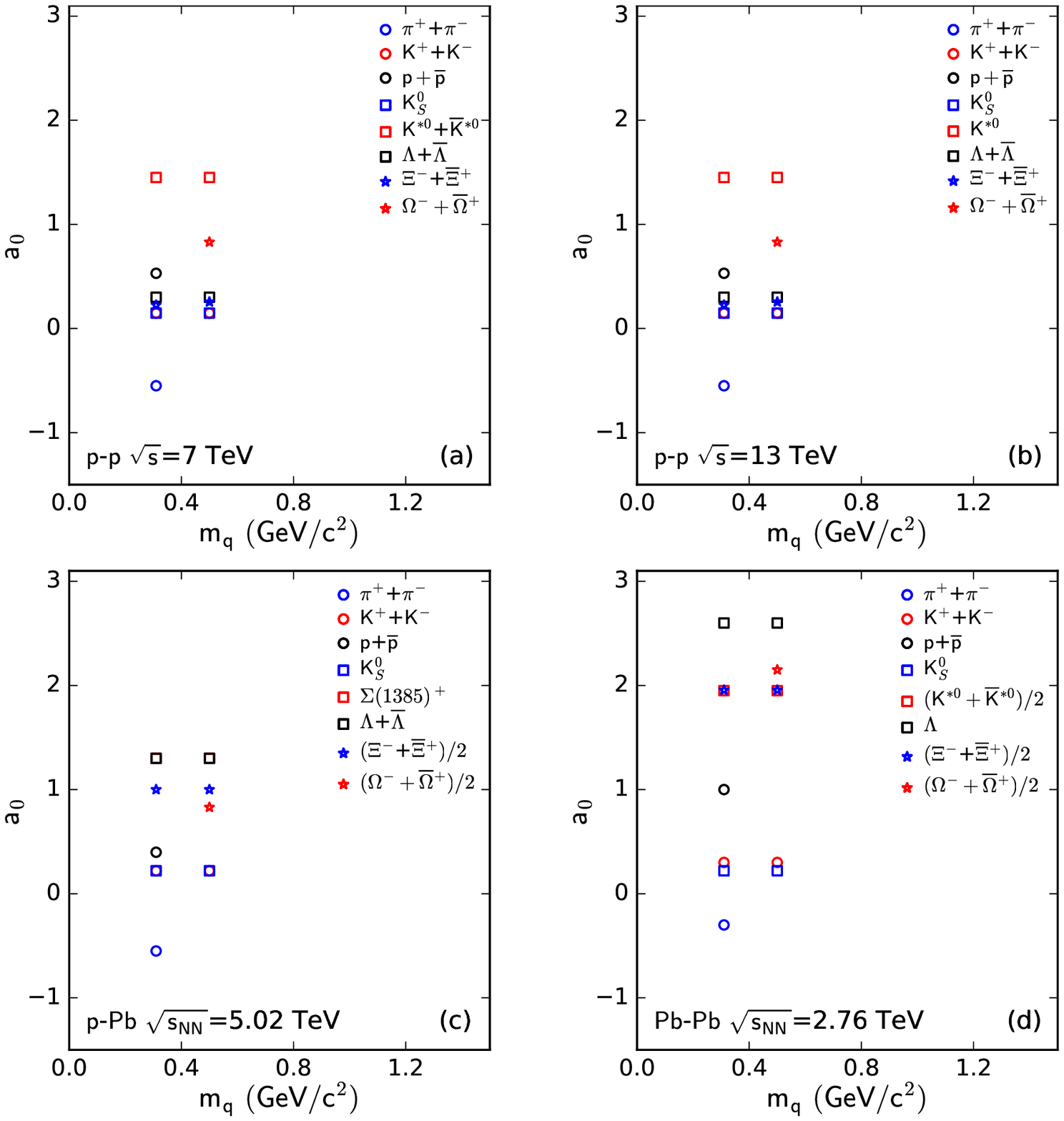}
\end{center}
\justifying\noindent {Fig. 10. Same as Figure 5, but showing the
dependence of $a_0$ on $m_q$.}
\end{figure*}

\begin{figure*}[htbp]
\begin{center}
\includegraphics[width=11.0cm]{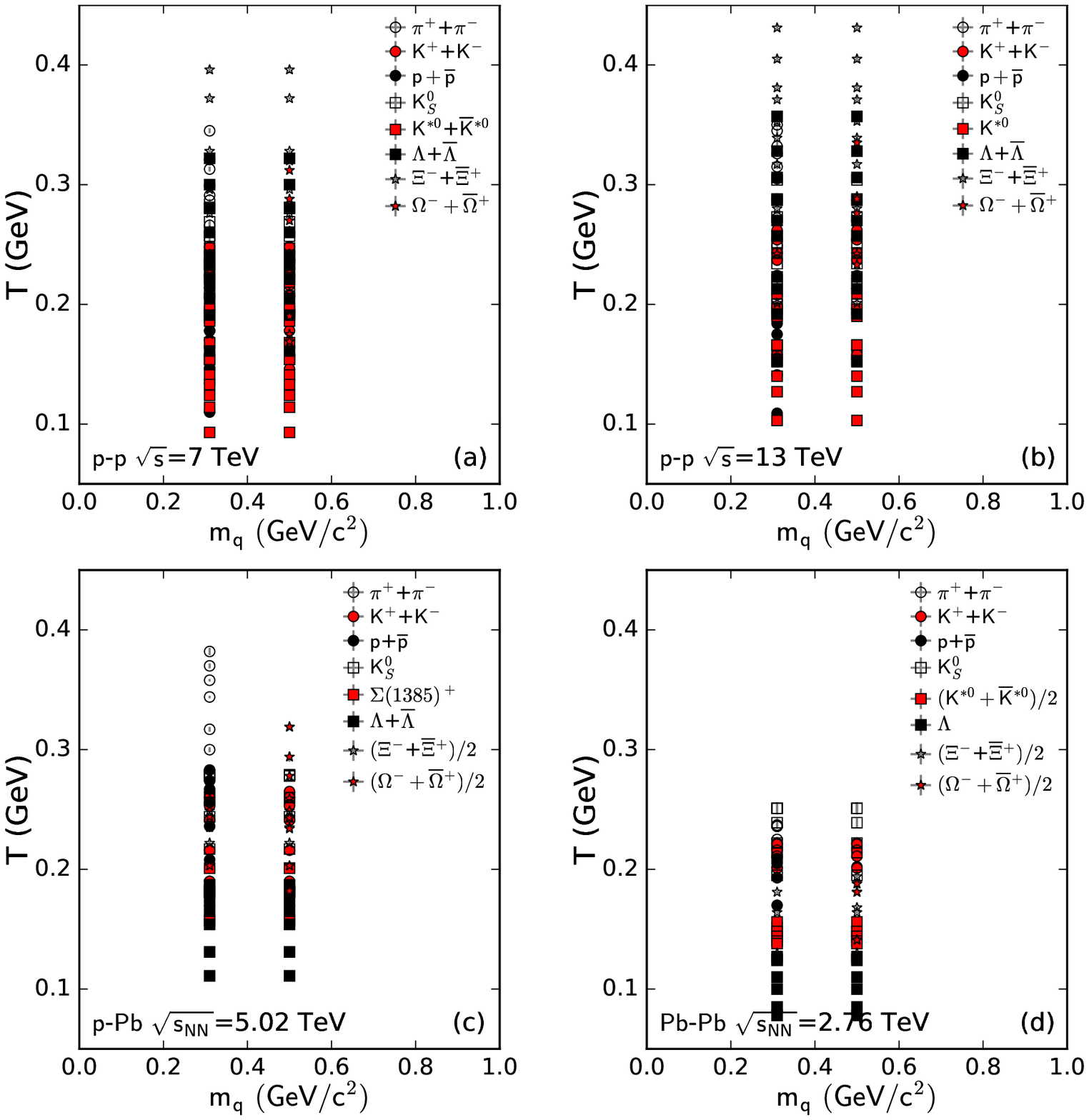}
\end{center}
\justifying\noindent {Fig. 11. Same as Figure 5, but showing the
dependence of $T$ on $m_q$.}
\end{figure*}

\begin{figure*}[htbp]
\begin{center}
\includegraphics[width=11.cm]{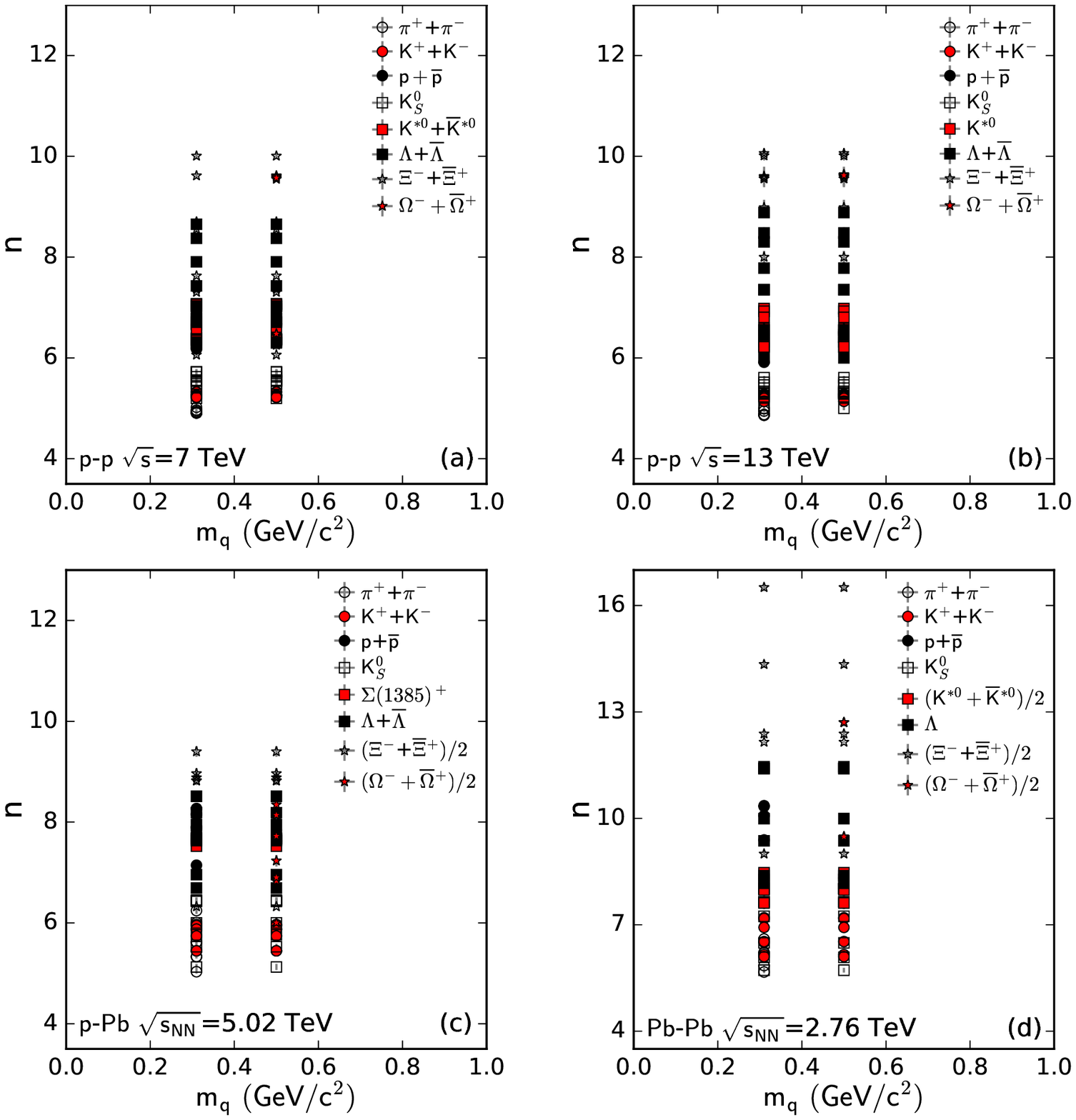}
\end{center}
\justifying\noindent {Fig. 12. Same as Figure 5, but showing the
dependence of $n$ on $m_q$}
\end{figure*}

In order to see more intuitively the dependence of parameters on
the centrality $C$ (the multiplicity is converted to centrality in
$pp$ collisions), rest mass $m_0$ of the hadron, and constituent
mass $m_q$ of the quark, we show the multi-factor dependent
parameters in Figures 5--12, where the values of parameters are
cited in Tables 1--4. Because of the revised index $a_0$ being
independent of the centrality, we have not shown the plot for the
centrality dependent $a_0$.

Figures 5 and 6 show the dependence of effective temperature $T$
and entropy-related index $n$ on centrality $C$ respectively.
Panels (a--d) are for $pp$ collisions at $\sqrt{s}=7$ TeV, $pp$
collisions at $\sqrt{s}=13$ TeV, $p$--Pb collisions at
$\sqrt{s_{\rm NN}}=5.02$ TeV, and Pb--Pb collisions at
$\sqrt{s_{\rm NN}}=2.76$ TeV, respectively. Different symbols
represent the parameters from the $p_T$ spectra of different
particles marked in the panels. One can see that with the decrease
of centrality from central to peripheral collisions, $T$ and $n$
decrease in most of the cases. Central collisions create higher
concentration of energies because of larger number of participants
and hence a higher system temperature, $T$ is expected. As a
consequence, this is expected to create a more thermalized system
leading to larger value of $n$ or smaller entropy index $q$, as
$n=1/(q-1)$. These results are consistent with the fact that more
central is the collision, the system has a higher tendency of
creating a high temperature thermalized system.

This result is consistent with our previous work~\cite{39} which
also shows that $T$ in central collisions is larger. In
Ref.~\cite{39} we have studied the light (also including
strangeness) and heavy particles produced in small system ($pp$,
$d$--Au, and $p$--Pb collisions) and large system (Au--Au and
Pb--Pb collisions) in the case of $\bm{p_{ti}}$ being parallel.
This result is also consistent with our another work published
recently~\cite{61}, which analyzes the kinetic freeze-out
temperature $T_0$ extracted from narrow $p_T$ spectra ($p_{T}<4.5$
GeV/$c$) of identified particles ($\pi^+$, $K^+$, $p$, $K^0_S$,
$\Lambda$, $\Xi$, $\Omega +\overline{\Omega}$) produced in
copper--copper (Cu--Cu), Au--Au, and Pb--Pb collisions by the
blast-wave model. That result shows larger $T_0$ in central
collisions.

Figures 7--9 show the dependences of parameters $a_0$, $T$, and
$n$ on the rest mass $m_0$ of particle, respectively. Same as
Figures 5 and 6, panels (a--d) are for $pp$ collisions at
$\sqrt{s}=7$ TeV, $pp$ collisions at  $\sqrt{s}$ =13 TeV, $p$--Pb
collisions at $\sqrt{s_{\rm NN}}=5.02$ TeV, and Pb--Pb collisions
at $\sqrt{s_{\rm NN}}=2.76$ TeV, respectively. Different symbols
represent the parameters from the $p_T$ spectra of different
particles marked in the panels. With the increase of $m_0$, one
can see that $a_0$ and $n$ increases significantly for the four
cases, and the tendency of $T$ is strange: $T$ decreases firstly
and then increases, the boundary is at around $m_0=1$ GeV/$c^2$.

The results of the $m_0$ dependent parameters are not
contradictory to our previous work~\cite{39} if we examine
minutely the parameter plots around $m_0=1$ GeV/$c^2$
there, though in which the analysis was done for a special case.
The amplitudes of the $m_0$ dependent parameters in Figures 7--9
are different for the four cases. These differences are explained
by different collision energies and system sizes. In particular,
in Pb--Pb collisions, multiple scattering of particles in hot and
dense matter affects the parameters. This multiple scattering also
reflects the shadowing effect in which the subsequent medium has
no chance to be collided due to the change of direction of motion of
the particles in the multiple scattering.

The dependences of parameters $a_0$, $T$, and $n$ on the
constituent mass $m_q$ of quark are given in Figures 10--12,
respectively. Same as Figures 5--9, panels (a--d) are for $pp$
collisions at $\sqrt{s}=7$ and 13  TeV, $p$--Pb collisions at
$\sqrt{s_{\rm NN}}=5.02$ TeV, and Pb--Pb collisions at
$\sqrt{s_{\rm NN}}=2.76$ TeV, respectively. Different symbols
represent the parameters from the $p_T$ spectra of different
particles marked in the panels. One can see nearly the same
parameters for $m_u$ ($m_d$) and $m_s$, and some differences for
the four collisions.

The present work on $m_q$ dependent parameters is not
contradictory to our previous work~\cite{39}, though two more
quarks (charm and bottom) are included there. The nearly the same
parameters are observed due to very small difference between $m_u$
($m_d$) and $m_s$ ($m_u=m_d=0.31$ GeV/$c^2$ and $m_s=0.5$
GeV/$c^2$~\cite{60a}). The differences in amplitudes for the four
collisions are explained by different collision energies and
system sizes, in particular by different system sizes, where the
influence of multiple scattering in large system is considerable,
as the explanations for Figures 7--9.

\subsection{Further discussion}

The values of revised index $a_0$ extracted from the $p_T$ spectra
of $\pi^++\pi^-$ are negative, which means the upward tendency of
the spectra in low-$p_T$ region~\cite{38}. This is contributed by
the resonance decays. The values of $a_0$ extracted from the
$p_T$ spectra of other particles are positive or even larger than
1, which means the downward tendency of the spectra in low-$p_T$
region. This is caused by the constraints in the production of
other particles. Larger constraint seems to be observed for
strange particles/quark. Due to the limitation of normalization,
any upward or downward tendency in low-$p_T$ region will cause the
shape change of the curve in intermediate- and high-$p_T$ regions.

The values of effective temperature $T$ extracted from the present
work contain both the contributions of thermal motion and
collective flow, which shows that $T$ is larger than 0.2 GeV even
0.3 GeV. To dissociate the two contributions and obtain the
kinetic freeze-out temperature $T_0$ and transverse flow velocity
$\beta_T$, one has to apply different methods which shows some
level of inconsistencies in some cases. These  inconsistencies mainly show in
different centrality and size dependent behaviour of $T_0$
and $\beta_T$. In our opinion, $T$ displays the coincident results
and should be paid more attention. To obtain $T_0$ and $\beta_T$,
a uniform method is needed at the first place. This issue is
beyond the focus of the present work. In all cases, the
entropy-related index $n>4$ which means that $q<1.25$, as expected
for high-energy collisions. This is a $q$ close to 1 and implies
that the system at the quark level is in approximate equilibrium
or local equilibrium.


As an extension of the special (parallel) case to general
(arbitrary) one, the results from our previous work~\cite{38,39}
are confirmed more broadly here. Combining with our previous work,
we may infer that the method used in the present work is suitable
for wide ranges of collision energy, system size, event
centrality, and particle species. Although the analytical
expression of $p_T$ distribution for the special case can be
obtained, we may use the Monte Carlo method to calculate the $p_T$
distribution for the general case.

We note that the fit for the $p_T$ spectra in Pb--Pb collisions is
not good, in particular for the spectra in high-$p_T$ region in
central collisions. This is explained by the influence of multiple
scattering of produced particles in hot and dense matter. To
include the influence of multiple scattering, except for the
superposition of the contributions of quarks with the TP-like form
of transverse momenta and isotropic azimuths, we may need another
function in the present framework. Or, we may use a two-component
function to fit the $p_T$ spectra in Pb--Pb collisions, where the
second component can describe the contribution of multiple
scattering.

The present work includes the contributions of soft excitations
and hard scattering processes together, without having a scope to
separate them in the ambient of the present analysis of identified
particle transverse momentum spectra using TP-like distribution
function. The possible departure of the fit from the data in
high-$p_T$ region in central Pb--Pb collisions could be due to
multiple scatterings and medium effects and in addition, a
thermalized medium being formed in these nuclear collisions.

\section{Summary and Conclusions}

The present analysis is summarized below with important
observations and conclusions.

(a) The transverse momentum, $p_T$ spectra of various light
hadrons (including some strange particles) produced in $pp$
collisions at $\sqrt{s}=7$ and 13 TeV, $p$--Pb collisions at
$\sqrt{s_{NN}}=5.02$ GeV, and Pb--Pb collisions at
$\sqrt{s_{NN}}=2.76$ TeV for different multiplicity or centrality
classes have been studied in the framework of a multi-source
thermal model at the quark level. The contribution of each
constituent quark to hadron's $p_T$ is assumed to obey the TP-like
function with isotropic azimuth. The calculations are performed by
Monte Carlo method and compared with the experimental data
measured by the ALICE Collaboration. A reasonably good description
of the spectra across various collision species and energies
available at the LHC is observed.

(b) With the decrease of final state multiplicity from central to
peripheral collisions, the revised index, $a_0$ being nearly
invariant, the effective temperature, $T$ and the entropy-related
index, $n$ decrease in most of the cases. With the increase of the
rest mass $m_0$ of the particles, $a_0$ and $n$ increases
significantly, and $T$ decreases first and then increases with the
boundary being at around 1 GeV/$c^2$. Higher collision energy is
deposited in central collisions, which results in larger $T$. The
system in central collisions is closer to the equilibrium, which
results in larger $n$ or smaller entropy index, $q$. Both the soft
excitation and hard scattering processes are considered uniformly.
The influence of multiple scattering in large system is
considerable.
\\
\\
\\
{\bf Acknowledgments}

The work of Pei-Pin Yang was supported by the China Scholarship
Council (Chinese Government Scholarship) under Grant No.
202008140170, the Shanxi Provincial Innovative Foundation for
Graduate Education under Grant No. 2019SY053, and the Innovative
Foundation for Graduate Education of Shanxi University. The work
of Shanxi Group was supported by the National Natural Science
Foundation of China under Grant Nos. 12047571, 11575103, and
11947418, the Scientific and Technological Innovation Programs of
Higher Education Institutions in Shanxi (STIP) under Grant No.
201802017, the Shanxi Provincial Natural Science Foundation under
Grant No. 201901D111043, and the Fund for Shanxi ``1331 Project"
Key Subjects Construction. Raghunath Sahoo acknowledges the
financial supports under the CERN Scientific Associateship, CERN,
Geneva, Switzerland and the financial grants under DAE-BRNS
Project No. 58/14/29/2019-BRNS of Government of India.
\\
\\
{\bf Author Contributions} All authors listed have made a
substantial, direct, and intellectual contribution to the work and
approved it for publication.
\\
\\
{\bf Data Availability Statement} This manuscript has no
associated data or the data will not be deposited. [Authors'
comment: The data used to support the findings of this study are
included within the article and are cited at relevant places
within the text as references.]
\\
\\
{\bf Ethical Approval} The authors declare that they are in
compliance with ethical standards regarding the content of this
paper.
\\
\\
{\bf Disclosure} The funding agencies have no role in the design
of the study; in the collection, analysis, or interpretation of
the data; in the writing of the manuscript, or in the decision to
publish the results.
\\
\\
{\bf Conflict of Interest} The authors declare that there are no
conflicts of interest regarding the publication of this paper.
\\
\\
\end{document}